\documentclass[12pt]{article}
\usepackage[top=1in, bottom=1.5in, left=1in, right=1in]{geometry} 
\usepackage[margin=1cm]{caption}

\usepackage{multirow}
\usepackage{natbib}
\usepackage{amsmath, bm}
\usepackage{amssymb}
\usepackage{graphicx}
\usepackage{multicol}
\usepackage{float}
\usepackage{hyperref}
\usepackage{color}
\usepackage{rotating}
\usepackage{subfig}
\usepackage{enumitem}
\usepackage{mathtools}
\usepackage{setspace}
\usepackage{verbatim}
\usepackage{soul}
\usepackage[normalem]{ulem}
\RequirePackage{etoolbox}
\RequirePackage{xcolor}
\definecolor{cb-blue}{RGB}{0, 109, 219}
\definecolor{cb-rose}{RGB}{255, 109, 182}

\DeclarePairedDelimiter\abs{\lvert}{\rvert}%
\makeatletter
\let\oldabs\abs
\def\abs{\@ifstar{\oldabs}{\oldabs*}}

\makeatletter
\def\namedlabel#1#2{\begingroup
    #2%
    \def\@currentlabel{#2}%
    \phantomsection\label{#1}\endgroup
}
\makeatother

\newtheorem{theorem}{Theorem}

\newtheorem{definition}{Definition}

\usepackage{A_command}

\begin{document}

\title{\large \bf Accuracy Gains from Privacy Amplification Through Sampling for Differential Privacy}

\author{
Jingchen Hu\footnote{Jingchen Hu is Assistant Professor at the Mathematics and Statistics Department, Vassar College, NY, USA. 
J\"org Drechsler is Distinguished Researcher at Institute for Employment Research, Nuremberg, Germany and Associate Research Professor in the Joint Program in Survey Methodology at the University of Maryland. 
Hang J. Kim is Associate Professor at the Division of Statistics and Data Science, University of Cincinnati, OH, USA.
*Address correspondence to Jingchen Hu, Mathematics and Statistics Department, Vassar College, 124 Raymond Ave, Box 27, Poughkeepsie, NY 12604, USA; E-mail: jihu@vassar.edu.}, 
J\"org Drechsler, 
and Hang J. Kim
}

\date{}

\maketitle

\abstract{Recent research in differential privacy demonstrated that (sub)sampling can amplify the level of protection. For example, for $\epsilon$-differential privacy and simple random sampling with sampling rate $r$, the actual privacy guarantee is approximately $r\epsilon$, if a value of $\epsilon$ is used to protect the output from the sample. 
In this paper, we study whether these amplification effects can be exploited systematically to improve the accuracy of the privatized estimate. Specifically, assuming the agency has information for the full population, we ask under which circumstances accuracy gains could be expected, if the privatized estimate would be computed on a random sample instead of the full population. We find that accuracy gains can be achieved for certain regimes. However, gains can typically only be expected, if the sensitivity of the output with respect to small changes in the database does not depend too strongly on the size of the database. 
We only focus on algorithms that achieve differential privacy by adding noise to the final output and illustrate the accuracy implications for two commonly used statistics: the mean and the median. We see our research as a first step towards understanding the conditions required for accuracy gains in practice and we hope that these findings will stimulate further research broadening the scope of differential privacy algorithms and outputs considered.} 




{\bf{keywords}}: Accuracy gains, Laplace mechanism, Privacy Amplification, Sensitivity 

\section*{Statement of Significance}

Our research studies whether amplification effects of sampling can be exploited systematically to improve the accuracy of the privatized estimate. Specifically, assuming the agency has information for the full population, we investigate under which circumstances accuracy gains could be expected, if the privatized estimate would be computed on a random sample instead of the full population. We find that accuracy gains can be achieved for certain regimes. However, gains can typically only be expected, if the sensitivity of the output with respect to small changes in the database does not depend too strongly on the size of the database. We see our research as a first step towards understanding the conditions required for accuracy gains in practice and we hope that these findings will stimulate further research broadening the scope of differential privacy algorithms and outputs considered.

\section{Introduction}
\label{intro}
In the era of digitisation, in which massive amounts of data are collected permanently about all aspects of our lives, ensuring confidentiality regarding the collected information becomes increasingly difficult. One strategy to address this challenge is to rely on formal privacy concepts, such as differential privacy \citep{Dwork:2006:CNS:2180286.2180305,dwork2008}, which offer mathematically provable privacy guarantees. Under the formal privacy framework, privacy is guaranteed irrespective of the assumptions regarding the additional information that might be available to an attacker trying to learn sensitive information.
The level of the privacy guarantee is expressed through \emph{privacy parameters}, for example, the parameter $\epsilon$ used in the original definition of differential privacy \citep{Dwork:2006:CNS:2180286.2180305}, where a smaller value of $\epsilon$ implies a higher level of protection.
However, an increased level of privacy protection comes at the price of reduced accuracy of the protected results. Thus, data providers have to carefully decide which value of $\epsilon$ to pick to ensure that the data still provide meaningful information while achieving a sufficient level of protection.

In recent years, some studies focused on quantifying the gains in the protection level, when computing the privatized output using only a sample of the original data.  
Intuitively, it seems reasonable that using only a sample of the original database will increase the level of protection, as the uncertainty about whether a specific target unit from the original database is included in the sample will increase. Statistical agencies have relied on this intuition for decades, for example, when releasing only a 1\% or 5\% sample of their census as public use microdata samples (PUMS), such as those available through IPUMS International. The rationale is that sampling itself offers so much protection that additional protection measures can be substantially reduced compared to releasing data for the full population. See, for example, \citet{drechsler2010,drechsler2012}, which illustrate this argument in the context of synthetic data \citep{Drechsler2011book}. 


Previous research \citep{KasiviswanathanLNRS11,LiQardajiSu2011,BunNSV15,WangLF16b,Abadi16,WangBK19,Balle:2018} found that the extra level of protection offered through sampling can indeed be formalized in the context of differential privacy. The privacy amplification effects from sampling are directly related to the sampling rate, that is, the smaller the sampling rate $n/N$, the higher the level of protection for the same value of $\epsilon$. 
Since most of the data collected by statistical agencies are based on sample surveys, these findings could be important for agencies willing to adopt the differential privacy approach as their data dissemination strategy. 


However, these encouraging findings 
only seem to hold for simple sampling designs, such as simple random sampling with or without replacement or Poisson sampling. Early research \citep{bun2020} indicates that privacy amplifications for commonly used sampling designs such as cluster sampling or stratified sampling are often small if they exist at all. The problem arises as these 
designs typically rely on some characteristics of the target units when setting up the sampling plan and thus the fact that a unit is included in the sample might leak additional information.


In this paper, we focus on another aspect of privacy amplification through sampling. 
Motivated by the fact that statistical agencies already rely on
sampling 
for data protection, we look at the problem from an accuracy perspective. 
Using the theoretical result that sampling allows for a larger $\epsilon$  
to achieve a fixed level of protection, we ask whether there are any regimes under which the accuracy of the reported results based on the sample will increase relative to the accuracy of the protected results computed on the full population. 
Obviously, there are two competing forces at play: reducing the sample size will increase the sampling error, while at the same time it will reduce the random error from data protection as larger values for $\epsilon$ can be used to protect the data, which implies less noise infusion. 
However, as we illustrate in the paper, there are other aspects that also need to be considered. For example, a key component of most differential privacy algorithms is the sensitivity of the statistic of interest with respect to small changes in the data. 
The specified statistic and the selected sensitivity concept (further discussed below) 
will influence how strongly the sensitivity depends on the sample size. 
Thus, when hoping for accuracy gains through sampling, it is important to pick algorithms whose sensitivity only weakly depends on the size of the data. We note that sampling might also be desirable for computational efficiency reasons in some cases. In such scenarios, trade-offs exist between computational complexity and accuracy of the results. Therefore, sampling might be considered useful even if some accuracy is sacrificed. As the trade-offs will be context specific in most cases, we will not consider the implications of computational efficiency in this paper.

The remainder of the paper is organized as follows. In Section \ref{dp}, we review some important concepts of differential privacy in the context of official statistics and provide some illustrations. Section \ref{amplification} provides a theorem regarding amplification effects of sampling, which is a rephrased version of previously published theorems looking at the problem from a survey statistics perspective. We then investigate accuracy gains from sampling in Section \ref{accuracy}, including illustrations for the mean and median. We end in Section \ref{conclusion} with some concluding remarks.

\section{Review of differential privacy}
\label{dp}

\subsection{$\epsilon$-Differential Privacy and $(\epsilon,\delta)$-Differential Privacy}
\label{dp:versions}


The original definition of $\epsilon$-differential privacy ($\epsilon$-DP) was given by \cite{Dwork:2006:CNS:2180286.2180305}. 
Here we directly quote the definition of $\epsilon$-DP from \citet{dwork2008}:
\begin{definition}
A randomized function $\mathcal{K}$ gives $\epsilon$-differential privacy if for all data sets $D$ and $D'$ differing on at most one element, and all $K \subset Range(\mathcal{K})$, 
\begin{equation} \label{eq:originalDP}
    \Pr[ \mathcal{K}( D ) \in K ] \le e^\epsilon \Pr[ \mathcal{K}( D' ) \in K ].
\end{equation}
The probability is taken over the coin tosses of $\mathcal{K}$. 
\end{definition}
In this definition, the condition is written as ``differing on \emph{at most} one element,'' but it suffices to consider the case that \emph{exactly} one element of each dataset differs because datasets that differ in no element always satisfy Equation \eqref{eq:originalDP} for any $\epsilon > 0$. Then, ``differing on exactly one element" covers two scenarios, typically known as bounded differential privacy and unbounded differential privacy. In the bounded case, the neighboring datasets $D$ and $D'$ are assumed to have the same number of elements, and exactly one element differs between the two datasets. The unbounded case assumes that $D'$ contains exactly one less (or one more) element than $D$ while all other elements are the same. In this paper, we discuss differential privacy under the bounded case.

To clarify the meaning of Definition \ref{eq:originalDP}, we adapt the definition to the context  in which a statistical agency aims at disseminating summary statistics or microdata to the public in a differentially private way.
Let $\mathcal{U}$ denote units in a population of interest and $\bmY_N = \{ y_i : i \in \mathcal{U} \}$ denote a set of variables $y_i$ for these population units, where the size of the population is given by $N = |\mathcal{U}|$. 
A statistical agency often disseminates summary statistics $g(\bmY_N) \in \Omega$. 
For example, the population mean and the population median are the solutions of the estimating equation $\sum_{i=1}^N U(\theta;y_i) = 0$ where the estimating functions are set as $U(\theta;y) = y - \theta$ and $U(\theta;y) = 0.5 - I[ y < \theta]$, respectively. 
If an $\epsilon$-DP mechanism is considered  for generating a privatized summary statistic, the randomized function $\mathcal{K}$ in the original definition is parameter-specific and can be written as a composite function $\mathcal{K}(\cdot) = \mathcal{A}(g(\cdot))$ where $\mathcal{A}$ is a randomized function $\mathcal{A}: \Omega \to \Omega$ representing the algorithm used for, for example by 
adding noise or smoothing.
Using this perspective, we rephrase the original $\epsilon$-DP definition as follows: 
\begin{definition}
A randomized function $\mathcal{A}: \Omega \to \Omega$ gives $\epsilon$-differential privacy for the population parameter $g(\bmY_N) \in \Omega$ if for a fixed value of $\epsilon > 0$ and for all $\omega \in \Omega$
\begin{equation} \label{eq:definition2}
\Pr \left[ \ \mathcal{A}(g(\bmY_N)) \le \omega \ \right] \le e^\epsilon \Pr \left[ \ \mathcal{A}(g(\bmY_N')) \le \omega \ \right] 
\end{equation}
for any $\bmY_N = \{ y_1, \ldots, y_N \}$ and $\bmY_N' = \{ y'_1, \ldots, y'_N \}$ differing on at most one pair of variables $y_i \neq y'_i$ for any $y_i, y'_i \in \mathcal{Y}$, where $\mathcal{Y}$ is the support of $\bmY$.
\end{definition}
This alternative definition emphasizes that the randomized function in Definition 1 is a composite function, composed of randomization for data protection and the analysis model or summary statistic of interest. The effect of the data protection varies with the statistic or the size of the protected data to be released to the public. We also hope that the alternative definition is more intuitive for the statistical community as it directly compares the CDFs of outputs of the randomized functions computed on two neighboring datasets $\bmY_N$ and $\bmY_N'$. We note that while this notation is most intuitive in the context of aggregate statistics, it also covers more general settings such as randomized response, in which the data are already altered at the time of collection. In this context $g(\cdot)$ would be the identity function and any analysis conducted on the collected data would be treated as post-processing.

This definition adopts the bounded version of differential privacy, which defines neighboring datasets as datasets having the same size but differing in one record. We believe this definition is more reasonable in settings where the data comprise the full target population. As the size of the population is fixed, it seems counter-intuitive to consider scenarios in which the population contains additional or fewer units. Arbitrary changes for one single record for a fixed-size database as considered in the bounded case seem more plausible in this context. While the size of the population is fixed, the data collected about this population do not need to be fixed.


Similarly, $(\epsilon,\delta)$-differential privacy ($(\epsilon,\delta)$-DP)  \citep{dwork2006our}, a relaxed version of $\epsilon$-DP,
is defined for finite population parameter dissemination as follows.
\begin{definition}
A randomized function $\mathcal{A}: \Omega \to \Omega$ gives $(\epsilon,\delta)$-differential privacy for the population parameter $g(\bmY_N)$ if for fixed values of $\epsilon > 0$ and $\delta \ge 0$ and for all $\omega \in \Omega$
\begin{equation} \label{eq:def:epsilon_delta}
\Pr \left[ \ \mathcal{A}(g(\bmY_N)) \le \omega \ \right] \le e^\epsilon \Pr \left[ \ \mathcal{A}(g(\bmY_N')) \le \omega \ \right] + \delta 
\end{equation}
for any $\bmY_N = \{ y_1, \ldots, y_N \}$ and $\bmY_N' = \{ y'_1, \ldots, y'_N \}$ differing on at most one pair of variables $y_i \neq y'_i$ for any $y_i, y'_i \in \mathcal{Y}$. 
\end{definition}
The definition of $\epsilon$-DP can be considered as a special case of the $(\epsilon,\delta)$-DP with $\delta$ set to zero. 

\subsection{Sensitivity: global, local, and smooth}
\label{dp:sensitivity}
 
A key quantity used in differential privacy algorithms, which governs how much a specific output needs to be perturbed to offer a target level of protection, is the sensitivity of the output. Intuitively, it measures the maximum change in the output given small changes in the database. Here, we review three definitions of sensitivity, each of which can be used with certain differential privacy mechanisms to achieve a given differential privacy definition for a specific statistic of interest. 

Using our notation introduced in Section \ref{dp:versions}, we formally define the \emph{global sensitivity} as follows:

\begin{definition}[Global Sensitivity]
For any  $g:\mathcal{Y}^N \to \Omega$, the global sensitivity is
\[
\Delta^G = \max_{\{ \forall \bm Y_N, \bm Y_N' \in \mathcal{Y}^N: \xi(\bm Y_N,\bm Y_N')\le1 \}}|g(\bmY_N) - g(\bmY_N')|,
\]
\end{definition}
where $\xi(\bmY_N,\bmY_N')\le1$ denotes that $\bmY_N$ and $\bmY_N'$ differ on at most one element.
In other words, the global sensitivity quantifies the maximum possible change of the statistic of interest if changing one record in the database. 
Intuitively, for a fixed privacy parameter, 
a larger value of $\Delta^G$ means that
$g$ is \emph{more sensitive} to changes in one record, 
so we need to introduce more uncertainty into the privatized output. 
In fact, a popular general approach for achieving $\epsilon$-DP 
is the Laplace mechanism (described in more detail below), which simply adds random noise to the output with variance proportional to $\Delta^G$. 

It is important to note that $\Delta^G$ is not tailored to a specific set of collected data, but applied to any possible dataset of size $N$. This implies that the global sensitivity can be large. To illustrate, we look at two common statistics that we will continue to use as illustrative examples throughout this paper: the mean and the median. 
Suppose that our data are bounded, that is, we can specify a range $R$ for which we know without looking at the data that no values will ever be outside this range.

To find the global sensitivity of the mean, we need to quantify the maximum possible change in the mean, if only a single record is changed. This maximum change occurs, if a value located at the lower bound of $R$ is changed to the upper bound of $R$. In this case, the change of the single value is $R$ and the change of the mean, that is, its global sensitivity, is $R/N$.
For the median assuming an odd number of observations, the global sensitivity is calculated based on the worst case scenario that the first  $(N+1)/2$ values including the median $m_N$ are located at the lower bound of $R$ and the remaining $N-(N+1)/2$ values are located at the upper bound of $R$. Under this scenario, if any value located at the lower bound is replaced by a value at the upper bound, the median switches from the lower bound to the upper bound. Thus, the maximum change of the median, that is, its global sensitivity, is the range $R$.
These examples highlight that the global sensitivity $\Delta^G$ 
depends on the functional form of $g$, but not on a specific dataset.

Unfortunately, when aiming at ensuring $\epsilon$-DP based on the global sensitivity, large values of $\Delta^G$ typically require introducing so much uncertainty that the privatized output will be useless. Also note that the global sensitivity for both of our illustrative statistics, mean and median, will be infinite if no natural bound can be specified.
For these reasons it seems natural to search for alternatives to the global sensitivity. One obvious candidate is the \emph{local sensitivity} defined as follows:
\begin{definition}[Local Sensitivity]
For any  $g:\mathcal{Y}^N \to \Omega$, the local sensitivity for the dataset $\bmY_N$ is
\[
\Delta^L(\bmY_N) = \max_{\{ \forall \bm Y_N' \in \mathcal{Y}^N: \xi(\bm Y_N,\bm Y_N')=1 \}}|g(\bmY_N) - g(\bmY_N')|.
\]
\end{definition}
The definition looks almost similar to the global sensitivity. However, the local sensitivity is a function of the input dataset $\bmY_N$
and quantifies the maximum possible change of the output if we change one record in the collected data. This value can be substantially smaller than  the global sensitivity. However, the local sensitivity can often not be used directly to achieve differential privacy, because the local sensitivity is data dependent. Because of this dependence, infusing noise proportional to the local sensitivity 
would already leak information regarding the underlying data \citep{DworkRoth2014}. To still be able to benefit from the substantial reduction in sensitivity compared to $\Delta^G$,  \citet{DworkLei2009STOC} developed the Propose-Test-Release approach, which first proposes a bound, $b$, on the local sensitivity, followed by a differentially private test to ensure the adequacy of the bound $b$. If the test is passed, one can use a differential privacy mechanism, such as the Laplace mechanism with a $b$-dependent scale parameter, to achieve $(\epsilon, \delta)$-DP for several commonly used statistics.



Coming back to our running examples, the local sensitivity of the mean is $R(\bmY_N)/N$, with $R(\bmY_N)=\max_{\{ \forall \bm Y_N \in \mathcal{Y}^N \}}\left( max(\mathcal{Y}) - min(\bmY_N), max(\bmY_N) - min(\mathcal{Y})\right)$. Assuming an odd number of observations, the local sensitivity of the median is \\
$\max \left( y_{ \left(\frac{N + 1}{2}+1\right) }- y_{\left(\frac{N + 1}{2}\right)},  y_{ \left(\frac{N + 1}{2}\right)} -  y_{\left(\frac{N + 1}{2}-1\right)}  \right)$, where $y_{(k)}$ denotes the $k$th smallest value in $\bmY_N$. There might be reductions when moving from $\Delta^G$ to $\Delta^L(\bmY_N)$ for the mean, if $R(\bmY_N)< R$. However, the effect will always be more substantial for the median, as we move from considering the entire (potentially unbounded) range $R$ to only considering the difference between the median and its two neighboring values.

Another approach to overcoming the challenge of local sensitivity being disclosive was proposed by \citet{NissimRaskhodnikovaSmith2007ACM}. The authors introduced the concept of \emph{smooth sensitivity}, which is an upper bound on the local sensitivity. Under the smooth sensitivity scheme, we still use the information from the collected data, but ensure that the information leakage is bounded by the defined privacy parameters. 

Here, we focus on an algorithm, which achieves $(\epsilon, \delta)$-DP for the median, as we will use this algorithm in the remainder of the paper.  
Interested readers are referred to \citet{NissimRaskhodnikovaSmith2007ACM} for a general definition of smooth sensitivity, as well as algorithms that can achieve $\epsilon$-DP for the median based on this concept. The following smooth sensitivity is used to guarantee $(\epsilon,\delta)$-DP
for the median based on a dataset with an odd number of observations:
\begin{equation}
    \Delta^\mathcal{S}(\bmY_N) = \textrm{max}_{k = 0, \cdots, N}\left(e^{-k\beta} \cdot \textrm{max}_{t = 0, \cdots, k+1}  \left( y_{\frac{N + 1}{2} + t} - y_{\frac{N + 1}{2} + t - k - 1} \right) \right),
    \label{eq:smooth_sensitivity_median}
\end{equation}
where $\beta = \epsilon / \left( 2 \log(2 / \delta) \right)$.
To achieve $(\epsilon,\delta)$-DP for the median, one can add Laplace noise to the confidential median with scale parameter proportional to $\Delta^\mathcal{S}(\bmY_N)$.

\subsection{The Laplace mechanism}
\label{dp:Laplace}

Among various differential privacy mechanisms, the Laplace mechanism is one of the most popular choices as it is simple to implement and works for any output for which the global sensitivity is known. With the Laplace mechanism the privacy guarantee defined in Equation \eqref{eq:definition2} is achieved by adding random Laplace noise to the output of interest. As the name suggests, the noise is drawn from a Laplace distribution with parameters based on the global sensitivity $\Delta^G$ and the privacy parameter $\epsilon$. 
\begin{definition}[Laplace mechanism]\label{def:Laplace}
For any  $g:\mathcal{Y}^N \to \Omega$, the Laplace mechanism is defined as
\[
\mathcal{A}(g(\bmY_N)) = 
g(\bmY_N) + \eta_N, \B \eta_N \sim \textrm{Laplace}(0, \Delta^G / \epsilon)
\]
where $\Delta^G$ is the global sensitivity of $g(\cdot)$.
\end{definition}

A proof that the Laplace mechanism satisfies $\epsilon$-DP can be found in \citet{DworkRoth2014}.
The variance of the specified Laplace distribution is $2\left(\Delta^G / \epsilon \right)^2$. 
Intuitively, this makes sense: the larger $\Delta^G$, the more sensitive the output, requiring more noise to be added to the output to achieve a target level of protection. Similarly, the smaller the privacy parameter $\epsilon$, the higher the target level of privacy, again requiring larger amounts of Laplace noise. We note that Definition \ref{def:Laplace} focuses on univariate outputs. When privatizing a multi-dimensional output, independent Laplace noise can be added to each component. 

\subsection{Illustrations}
\label{dp:illustrations}

\begin{figure}[t]
    \centering
    \includegraphics[width=0.45\textwidth]{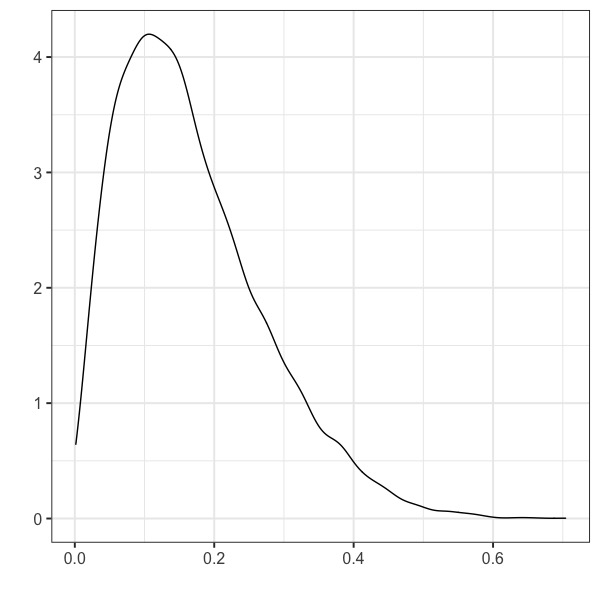}
    \includegraphics[width=0.45\textwidth]{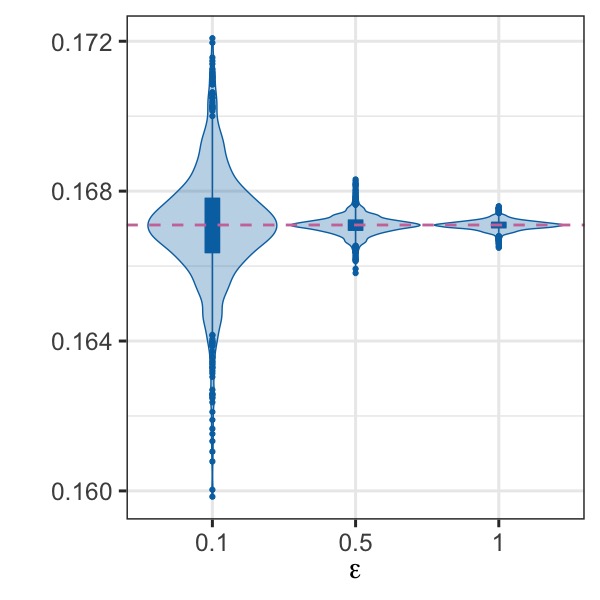}
    \caption{Density plot of the simulated population of size $N = 10,001$ generated from Beta(2, 10) (left panel) and violin plots of the privatized means via the Laplace mechanism for $\epsilon \in \{0.1, 0.5, 1\}$ 
    based on 1,000 realizations (right panel). The true  mean is plotted as the horizontal dashed line for comparison. 
    }
    \label{fig:beta2and10_eDP_pop_all}
\end{figure}

We close this section with a few illustrations of the impact of the sensitivity on the accuracy of the privatized mean and median as the outputs of interest. We simulate a database comprising $N=10,001$ records from a Beta(2,10) distribution which has a range of $R=1$. Hence, the global sensitivity of the mean is $R/N=1/10,001$ and that of the median is $R=1$. The empirical density of the generated data is shown in Figure \ref{fig:beta2and10_eDP_pop_all}.

To illustrate the amount of noise that needs to be added to achieve 
a target level of privacy, 
we repeatedly ($T= 1,000$ times) apply the Laplace mechanism for different levels of $\epsilon$
$\in\{0.1,0.5,1\}$ to obtain privatized means and medians. We use violin plots to visualize our results.

The right panel of Figure \ref{fig:beta2and10_eDP_pop_all}
shows the distribution of the privatized means from the repeated simulations.
Not surprisingly the violin plots become more concentrated around the mean of the non-privatized data $\bar{y}_N$ (the horizontal dashed line) as $\epsilon$ 
increases. In other words, lower privacy protection levels result in privatized means with higher accuracy. The relatively small global sensitivity value scaled to the inverse of the population size, $1 / N$, helps produce privatized means with relatively high utility overall, even if the privacy parameter $\epsilon$
is small. 

\begin{figure}[t]
    \centering \includegraphics[width=0.45\textwidth]{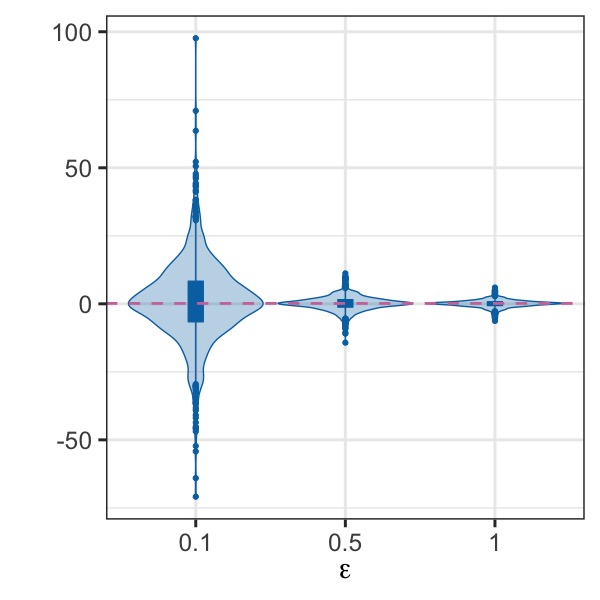}
    \includegraphics[width=0.45\textwidth]{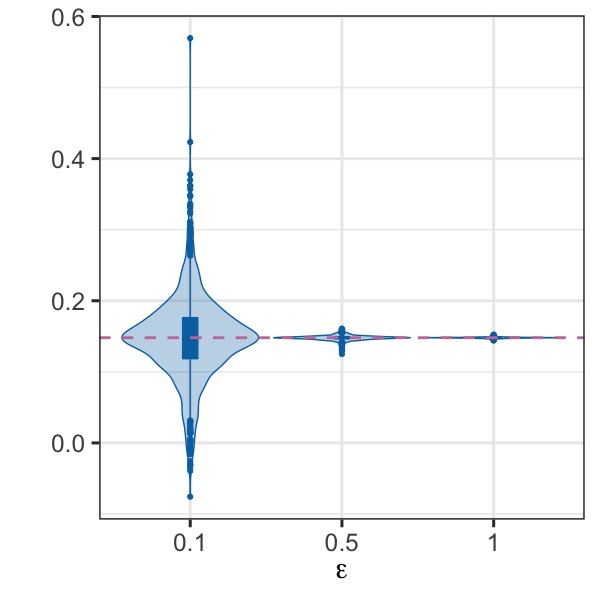}
    \caption{Violin plots of the privatized medians using global sensitivity (left panel) and smooth sensitivity (right panel) for $\epsilon \in \{0.1, 0.5, 1\}$ (and $\delta = 4.9995 \times 10^{5}$ for smooth sensitivity) based on 1,000 realizations. The simulated population of size $N = 10,001$ is generated from Beta(2, 10). The median of the original data is plotted as the horizontal dashed line.
    }
    \label{fig:PopulationMedian_beta2and10_eDP_pop_all}
\end{figure}

This does not hold if the Laplace mechanism
is used for the median. As can be seen in 
the left panel of Figure \ref{fig:PopulationMedian_beta2and10_eDP_pop_all}, the amount of noise that needs to be added is so large especially for small $\epsilon$
that the privatized medians become useless: when $\epsilon = 0.1$, 
the privatized median ranges from about -50 to 100, for data bounded in [0, 1]. 

To avoid excessive amounts of noise, 
one may consider protecting the median based on its smooth sensitivity
as discussed in Section \ref{dp:sensitivity}. Results based on the smooth sensitivity with $\delta = 1/(2N) = 4.9995 \times 10^{-5}$
are shown in 
the right panel of Figure \ref{fig:PopulationMedian_beta2and10_eDP_pop_all}.
The reduction in the variability of the privatized estimate is striking. Even for $\epsilon=0.1$, almost all estimates are within the support of the data and the level of uncertainty in the noisy estimates seems to be acceptable for practical purposes for $\epsilon=0.5$ or higher. 

Our illustrations demonstrate that for 
privatized medians, using the smooth sensitivity instead of the global sensitivity can offer large gains in accuracy if the global sensitivity is high. Of course, it must be noted that the price for these gains is the move from strict $\epsilon$-DP to 
$(\epsilon,\delta)$-DP.

\section{Amplification effects of sampling}
\label{amplification}


Previous research showed that sampling can \emph{amplify} the level of the differential privacy guarantee  \citep{LiQardajiSu2011,Balle:2018}. 
In this section, we will reiterate some of the previous findings using notation from the survey sampling literature. Let $\mathcal{S}$ denote a probability sampling mechanism that assigns inclusion indicators $Z_i$ to all population units in $\mathcal{U}=\{1,\ldots,N\}$ such that $Z_i=1$ if unit $i$ is selected in the sample and $Z_i=0$ otherwise. Given the selected sample, a finite population parameter of interest $g(\bmY_N)$ is estimated by the sample estimate $g(\mathcal{S}(\bmY_N))$.
The following theorem introduced in \cite{LiQardajiSu2011} and \cite{Balle:2018}
shows the \emph{amplification effects} of sampling if we consider 
drawing a sample of size $n$ using simple random sampling without replacement.
\begin{theorem} \label{eq:amp_theorem}
Suppose that a randomized function $\mathcal{A}$ gives $(\epsilon,\delta)$-DP for the population parameter $g(\bmY_N)$.
Then, $\mathcal{A}$ satisfies $(\epsilon_n,\delta_n)-$DP for the estimator $g(\mathcal{S}(\bmY_N))$ where $\epsilon_n = \log( 1 + n/N ( e^\epsilon - 1 ))$ and $\delta_n = (n/N)\delta$, that is,
for fixed values of $\epsilon > 0$ and $\delta \ge 0$ and for all $\omega \in \Omega$
$$
\Pr \left[ \ \mathcal{A}(g(\mathcal{S}(\bmY_N))) \le \omega \ \right] \le \left( 1 + \frac{n}{N} ( e^\epsilon - 1 )) \right) \Pr \left[ \ \mathcal{A}(g(S(\bmY_N'))) \le \omega \ \right] + \frac{n}{N} \delta  
$$
for any $\bmY_N = \{ y_1, \ldots, y_N \}$ and $\bmY_N' = \{ y'_1, \ldots, y'_N \}$ differing on at most one record $y_i \neq y'_i$ with any $y_i, y'_i \in \mathcal{Y}$. 
\end{theorem}

The proof of Theorem 1 under the bounded version of differential privacy is provided in Appendix A. Our proof differs from the proofs given in previous papers. While \cite{LiQardajiSu2011} only focus on the unbounded case, \cite{Balle:2018} use the ideas of coupling and total variation distance for their proof. We believe that our proof is more intuitive for the statistical community. 





Theorem \ref{eq:amp_theorem} shows that the privacy parameters for releasing  $\mathcal{A}(g(\mathcal{S}(\bmY_N)))$ are smaller than those for $\mathcal{A}(g(\bmY_N))$, that is, the simple random sampling scheme $S$ \emph{amplifies} the data protection level. 
This result can be useful for two data collection and dissemination scenarios. 
The most obvious application would be in the context of sample surveys in which the statistical agency collects a survey sample, $\mathcal{S}(\bmY_N)$, to obtain a sample estimate of the finite population parameter, $g(\mathcal{S}(\bmY_N))$. If the agency aims at a target level of protection specified through fixed values $(\epsilon,\delta)$, the actual values of the privacy parameters will be calculated based on $g(\mathcal{S}(\bmY_N))$, not $g(\bmY_N)$.
These values will generally be larger than $(\epsilon,\delta)$. 
Specifically, for simple random sampling without replacement, Theorem \ref{eq:amp_theorem} shows that privatizing the sample estimates with $\epsilon_n = \log(1 + (N/n)(e^\epsilon - 1))$ and $\delta_n = (N/n)\delta$ suffices to ensure the target level of the privacy guarantee, $(\epsilon,\delta)$. This would be attractive as less noise would need to be infused to achieve a target level of protection. 

\begin{figure}[t]
\centering
\includegraphics[width=0.5\textwidth]{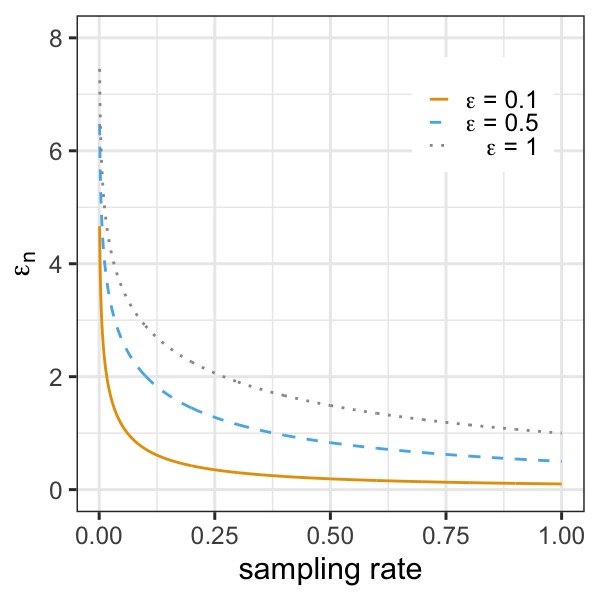}
\caption{
Amplification effects for
different values of the target privacy parameter
($\epsilon \in \{0.1, 0.5, 1\}$) 
and sampling rate $n/N \in (0, 1]$.}
\label{fig:epsilon_s_sampling_rate}
\end{figure}

Figure \ref{fig:epsilon_s_sampling_rate} illustrates that the amplification effects for $\epsilon$ can be substantial for small sampling rates. For example, to achieve $\epsilon=1$ based on a 1\% simple random sample it would be sufficient to use $\epsilon_n=5.15$ when protecting the output based on the sample. Also note that the amplification effects of $\delta$ are always linear in the sampling rate.

However, as noted in Section \ref{intro}, more complex designs than simple random sampling will typically be used in practice and it is by no means obvious that similar amplification results are achievable for these sampling designs. In fact, early research \citep{bun2020} indicates that amplification might be difficult to achieve for stratified and cluster sampling designs. 

An alternative scenario of using the amplification effects of sampling, which will be the focus of this paper, arises, if the statistical agency collects data for the entire population $\bmY_N$, either through administrative processes or by conducting a census, and considers sampling from these data for confidentiality reasons relying on the amplification effects. However, this approach would come at the expense of introducing sampling variance and possibly an increased sensitivity of the estimate to changes of the input data. Under this scenario, the focus shifts from understanding the privacy implications, which are directly quantifiable based on Theorem \ref{eq:amp_theorem} assuming simple random sampling, to measuring the impacts on accuracy. For a target level of protection specified through ($\epsilon$, $\delta$), the agency can search for the optimal sampling rate that minimizes the overall variance of the final estimate. To obtain this rate, the accuracy gains from the amplification effects need to be balanced against the increased sampling variance and the potential increase of the sensitivity of the desired output. 

\section{Accuracy gains from sampling}
\label{accuracy}
The variance of the privatized output $\mathcal{A}(g(\mathcal{S}(\bmY_N)))$ will always depend on two components: the sampling variance and the variance from noise infusion. When focusing on the class of differential privacy algorithms based on noise addition such as the Laplace mechanism, 
accuracy gains can only be expected
if the reduction of the variance from noise infusion is larger than the increase in sampling variance.
While there is an inverse proportional relationship between the sample size and the sampling variance for most 
estimators 
(if we ignore the finite population correction factor), this relationship is more nuanced for the noise component. Figure \ref{fig:epsilon_s_sampling_rate} already reveals the non-linear relationship between the sampling rate and the privacy amplification. This non-linear relationship implies that we will likely only see accuracy gains for small sampling rates. 

For many outputs, this discussion is further complicated by the fact that the sensitivity will depend on the size of the database. Thus, the gains in accuracy from increasing $\epsilon_n$ will be counter balanced by the increase in sensitivity, which intuitively will always require more noise infusion for any differential privacy algorithm.

So the natural question arises: Under which circumstances can accuracy gains be expected? In this section, we will discuss some aspects that need to be considered in this context focusing primarily on the Laplace mechanism. We also illustrate through simple examples that it is critical to ensure that the sensitivity of the output does not depend too strongly on the size of the database, when hoping for accuracy gains. 

\subsection{Some results for the Laplace mechanism}
\label{amplification:variance}

Let $V_n$ denote the variance of the privatized estimate $\mathcal{A}(g(\mathcal{S}(\bmY_N))))$ and let $V_N$ denote the variance of the privatized 
population
parameter $\mathcal{A}(g(\bmY_N)))$.
Using the Laplace mechanism we have $\mathcal{A}(g(\mathcal{S}(\bmY_N))) = g(\mathcal{S}(\bmY_N)) + \eta_n$, where $\eta_n \sim \textrm{Laplace}(0, \Delta_n^{G} / \epsilon_n)$ and $\Delta_n^G$ is the global sensitivity of the output in the sample. The variance of the privatized estimate 
is given by
\[
    V_n
    = \Var(g(\mathcal{S}(\bmY_N))) +  \Var(\eta_n) = \Var(g(\mathcal{S}(\bmY_N))) +  \frac{2 (\Delta_n^{G})^2 }{\epsilon_n^2}, \label{eq:variance_sum}
\]
where $\textrm{Var}(g(\mathcal{S}(\bmY_N)))$ is the sampling variance of the estimate and $2 (\Delta_n^{G} / \epsilon_n)^2$ is the variance of the Laplace noise.

Making general statements when to expect $V_n \leq V_N$ is difficult, as both the sampling variance as well as the variance from noise infusion will depend on the functional form of $g$. However, $V_n \leq V_N$ implies that 
the sampling variance always needs to be smaller than $V_N$  (as Var$(\eta_n)$ $> 0$), that is, the sampling variance always needs to be smaller than the variance of the noise added to protect the population parameter. In many cases, this might already imply that the sampling rate cannot be too small to avoid large sampling variances. 
However, large sampling rates imply that the variance from noise infusion will not be substantially reduced compared to the population. Thus, it is important to also account for the noise infusion for the sample estimate.

Tighter bounds for the sampling variance which also account for 
Var$(\eta_n)$ can be found, if we notice that it is reasonable to assume that the global sensitivity $\Delta^G$ will typically only increase if the size of the database is decreasing. Thus, the best case scenario from an accuracy gains perspective would be that the sensitivity is \emph{independent} of the sampling rate, that is $\Delta^G_n = \Delta^G_N$, as this would maximize $(V_N - V_n)$. Under this scenario, we can rewrite the bound of the sampling variance as follows:
\begin{eqnarray}\label{eq:samp_var_bound}
\textrm{Var}(g(\mathcal{S}(\bmY_N)))&\leq& V_N-Var(\eta_n) 
\ \leq \ \left(1-\frac{Var(\eta_n)}{V_N}\right)V_N\\
\nonumber&\leq&\left(1-\frac{2(\Delta^G_n/\epsilon_n)^2}{2(\Delta^G_N/\epsilon)^2}\right)V_N
\ \leq \ \left(1-\frac{\epsilon^2}{\epsilon_n^2}\right)V_N \\
\nonumber& \leq & q \ V_N.
\end{eqnarray}
\begin{figure}[t]
\centering
\includegraphics[width=0.6\textwidth]{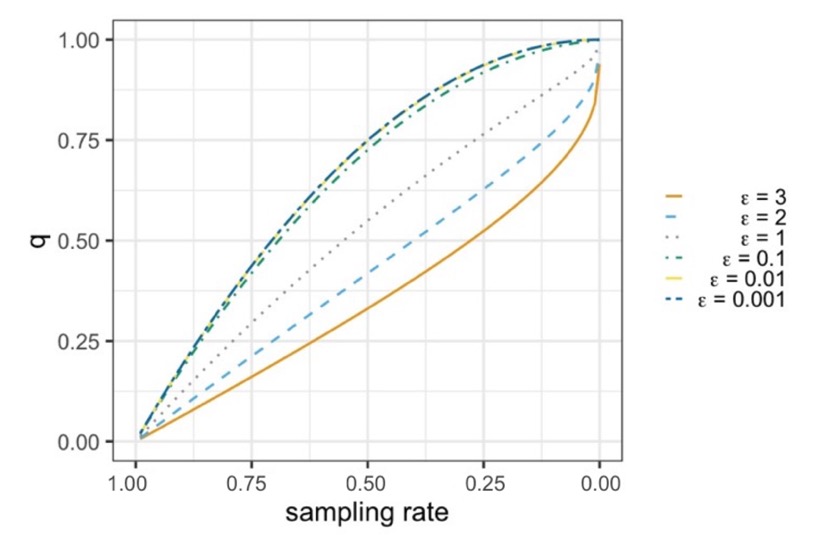}
\caption{Relationship between $q$ and the sampling rate for various levels of $\epsilon$.}
\label{fig:q_vs_rate}
\end{figure}
Figure \ref{fig:q_vs_rate} illustrates the relationship between the sampling rate and $q$ for various levels of $\epsilon$. The figure reveals several important findings. 
For $\epsilon>1$, $q$ increases at a less than linear rate as sampling rates decrease (except for very small sampling rates). 
For example, the solid curve in the figure suggests that 
the sampling rate would need to be less than $16.77\%$ to still see accuracy gains for $\epsilon=3$ if the sampling variance is $60\%$ of $V_N$. 
For values of $\epsilon<1$ the relationship is reversed. The parameter $q$ increases quickly when sampling rates are large but the rate of increase slows down with decreasing sampling rates. Thus, the same benchmark of $0.6 V_N$ could be achieved with a sampling rate of $61.4\%$ for $\epsilon=0.1$. Given that $V_N$ with $\epsilon=0.1$ will 
be 900 times larger than $V_N$ based on $\epsilon=3$, it is obvious that it will be much more likely to find regimes that offer accuracy gains for smaller values of $\epsilon$. However, Figure \ref{fig:q_vs_rate} also reveals that there is a bound to this effect. For $\epsilon$ values less than 0.1, the curve hardly changes, that is, for a fixed sampling rate, the value of $q$ would be almost the same for $\epsilon=0.01$ and $\epsilon=0.001$. This effect can be explained if we notice that for small $x$ it holds that $e^{x}\approx 1+x$, which implies that 
\begin{eqnarray*}
q&=&\left(1-\frac{\epsilon^2}{\epsilon_n^2}\right)
\ = \ \left(1-\frac{\epsilon^2}{(\log(1 + (N/n)(e^\epsilon - 1)))^2}\right) \\
& \approx & \left(1-\frac{\epsilon^2}{(\log(1 + (N/n)(1+\epsilon-1)))^2}\right)
\ = \ \left(1-\frac{\epsilon^2}{(\log(1 + (N/n)\epsilon))^2}\right) \\
& \approx & \left(1-\frac{\epsilon^2}{(\log(e^{N/n\epsilon}))^2}\right)
\ = \ \left(1-\frac{\epsilon^2}{(N/n\epsilon)^2}\right) \\ 
&=& 1-\left(\frac{n}{N}\right)^2.
\end{eqnarray*}
Thus, $q$ will only depend quadratically on the sampling rate, but will no longer depend on $\epsilon$
. Of course, it will still remain true that reducing $\epsilon$ will increase $V_N$. Thus, further decreasing $\epsilon$ will still make it more plausible to find regimes which offer accuracy gains.

However, these results only give an upper bound for the sampling variance, assuming that the sensitivity of the output is independent of the sample size. The bound will be tighter if the sensitivity depends on the size of the database since 
Var$(\eta_n)/V_N$ will get larger. A more general finding can be obtained, if we restate Equation (\ref{eq:samp_var_bound}) as a negative result. Noting that $V_N=2 (\Delta_N^{G} / \epsilon)^2$, we find that we can never expect to see accuracy gains whenever
\[
\textrm{Var}(g(\mathcal{S}(\bmY_N)))\geq 2 (\Delta_N^{G})^2\left(\frac{1}{\epsilon^{2}}-\frac{1}{\epsilon_n^{2}}\right).
\]
This result holds for any $g$ as long as the Laplace mechanism is used to protect the output.

To get a better understanding under which circumstances accuracy gains could be expected in practice, we evaluate the amplification results for two specific examples, assuming
interest lies in privatizing the mean and the median of the population.

\subsection{Accuracy gains for the population mean}
\label{amplification:mean}

Let $\bar{y}_N$ be the population mean, $S_N^2 = \sum_{i=1}^{N}(y_i - \bar{y}_N)^2 / (N-1)$ be the population variance, and $R$ be a known upper bound for the range of $\mathcal{Y}$ which is not data-specific. For this illustration, we assume the statistical agency relies on the Laplace mechanism with global sensitivity $\Delta^G = R/N$. 
Then, the variance of the privatized population parameter
$\mathcal{A}(g(\bmY_N))$ will be
\begin{equation}
    V_N =
    \textrm{Var}\left(\mathcal{A}(\bar{y}_N)\right) = 2 \left(\frac{R}{\epsilon N}\right)^2,
    \label{eq:pop_mean_var_pop}
\end{equation}
which accounts only for the Laplace noise component because the sampling variance is zero.


If the agency draws a simple random sample without replacement from the population and releases a privatized sample mean $\mathcal{A}(\bar{y}_n)$,  
its global sensitivity is  $\Delta_n^G = R / n$. 
The amplified privacy parameter is given by $\epsilon_n = \log(1 + (N/n)(e^\epsilon - 1))$, which implies that the Laplace noise has variance $\Var(\eta_n) = 2 R^2 / (\epsilon_n n )^2$. Therefore, the total variance of the privatized sample mean is
\begin{equation}
    V_n =
    \textrm{Var}\left(\mathcal{A}(\bar{y}_n)\right) = 
    \left(1 - \frac{n}{N}\right)\frac{S_N^2}{n} +  2 \left(\frac{R}{\epsilon_n n}\right)^2,
    \label{eq:pop_mean_var_sample}
\end{equation}
where the first term is the sampling variance of the sample mean and the second term is the Laplace noise component.

Making general statements regarding the accuracy gains or losses when using $\mathcal{A}(\bar{y}_n)$ instead of $\mathcal{A}(\bar{y}_N)$ is again difficult, as the results will depend on the population variance $S_N^2$ and the range $R$. However, we can gain some insights by only comparing the noise component of the two algorithms. 
Specifically, the ratio of the noise component for the population parameter to the noise component of the sample estimate is given by:
\begin{equation}
    r \left(\epsilon, \frac{n}{N} \right) = \frac{2 \left(\frac{R}{\epsilon N}\right)^2}{2 \left(\frac{R}{\epsilon_n n}\right)^2} 
    = \left( \frac{n}{N} \frac{\log(1 + \frac{N}{n}(e^\epsilon - 1))}\epsilon \right)^2. 
    \label{eq:pop_ratio}
\end{equation}

Note that as long as this ratio is smaller than one, we cannot hope for any accuracy gains when releasing the privatized sample mean, $\mathcal{A}(\bar{y}_n)$, as its variance from noise addition would be larger than that in the population. Thus, we would not see any accuracy gains even if we ignore the sampling variance. We also note that the ratio approaches 1, if we let $(N/n)\epsilon\rightarrow0$, since at least asymptotically it holds that
\[
\log\left(1 + \frac{N}{n}(e^\epsilon - 1)\right)\approx \log \left(1 + \frac{N}{n}(1+\epsilon - 1)\right)=\log\left(1 + \frac{N}{n}\epsilon\right)\approx \log(e^{(N/n)\epsilon})=\frac{N}{n}\epsilon.
\]
Thus, as $(N/n)\epsilon\rightarrow0$, 
the amount of noise that needs to be added will always be the same irrespective of the sampling rate, and accuracy gains would be impossible.

\begin{figure}[t]
    \centering
     \includegraphics[width=0.45\textwidth]{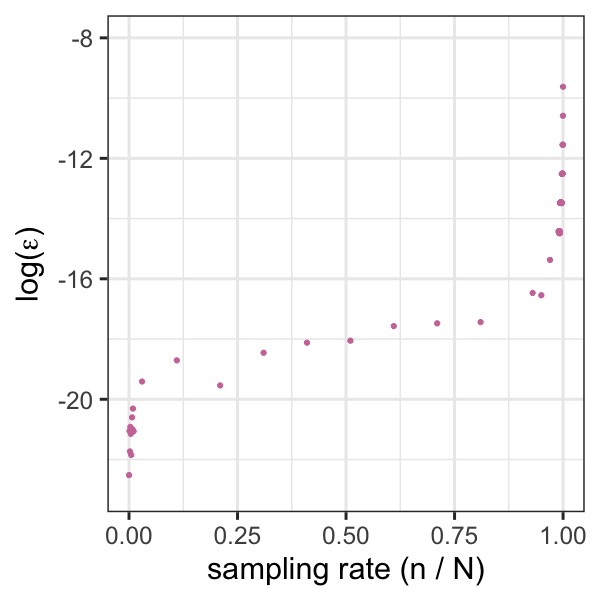}
     \includegraphics[width=0.45\textwidth]{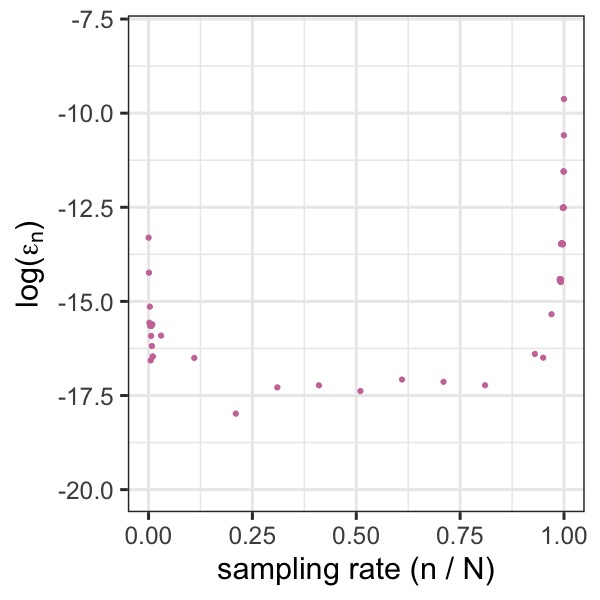}
        \caption{Dot plots of 
        the log of the target privacy parameter $\epsilon$ (left panel) and the log of amplified $\epsilon_n$ used for Laplace noise generation (right panel) for specific sampling rates $n/N$. The depicted values ensure that the amount of noise added to $\bar{y}_N$ matches the amount of noise added to $\bar{y}_n$, that is, $r(\epsilon, n / N) = 1$.}
    \label{fig:mean_r}
\end{figure}

Still, given that the ratio $r(\epsilon, n/N)$ is only a function of $\epsilon$ and the sampling rate $n/N$, we can evaluate analytically, that is, without relying on asymptotics,  which combinations of sampling rates and $\epsilon$ would lead to $r(\epsilon, n/N)=1$.
This relationship is depicted in the left panel of Figure \ref{fig:mean_r}. Note that $r(\epsilon, n/N)>1$ (implying that $Var(\eta_n)<V_N$) could only be achieved for $\epsilon$ values below the depicted curve.
The figure reveals that the required value of $\epsilon$ decreases with the sampling rate. Even more importantly, we find that the required level of $\epsilon$ needs to be extremely small to achieve $r(\epsilon, n/N)\leq1$ for all sampling rates. The values of $\epsilon$ are depicted on the log scale and a value of $\log(\epsilon)=-8$ already corresponds to $\epsilon\approx 0.0003$ whereas a value of $\log(\epsilon)=-20$ implies that $\epsilon\approx2 \times 10^{-9}$.

However, since smaller sampling rates imply larger amplifications effects, it might arguably be more relevant to focus on the relationship between the sampling rate and $\epsilon_n$, which represents the actual value of the privacy parameter that is used for a given sampling rate. This relationship is depicted in the right panel of Figure \ref{fig:mean_r}. We see that because of the amplification effects, the value of $\epsilon_n$ starts to increase again once we reach sampling rates less than 10\%. Still, even for a sampling rate of 0.0001 (the left most dot in the figure), $\epsilon_n<
1.66 \times 10^{-6}$ would be required to ensure that the ratio is less than one. 
It seems unlikely that statistical agencies will ever use such small values for $\epsilon_n$ in practice. While the privacy guarantees would obviously be high, the amount of noise that would need to added would be so large that any results obtained would no longer be meaningful. 
We also emphasize that $r(\epsilon, n/N)<1$ only ensures that the variance of the noise component will be smaller for the sample mean than for the population mean. These evaluations still completely ignore the sampling variance. Thus, it seems unlikely that the accuracy of the protected sample mean can ever be higher than the accuracy of the protected population mean.



These results illustrate that accuracy gains are difficult to achieve if the sensitivity strongly depends on the sample size. Thus, to even hope for accuracy gains form sampling, it is fundamental to use algorithms which show little dependence between the sensitivity and the sample size. In the next sections we illustrate that accuracy gains are indeed possible for certain combinations of outputs and algorithms.

\subsection{Accuracy gains for the population median}
\label{amplification:median}

The median is another popular statistic to offer insights regarding the location of the distribution. For skewed distributions such as income, the median is preferred to the mean because of its robustness to rare events in the tail. For data points $y_1, \cdots, y_N$ in the finite population $\bmY_N$ bounded within range $R$, let $y_{(k)}$ denote the $k$-th smallest value of $\bmY_N$. 
Let $m_N$ be the population median, which is $y_{\left(\frac{N + 1}{2}\right)}$ when $N$ is odd and $( y_{\left(\frac{N}{2} \right)} + y_{\left(\frac{N}{2}+1\right)}) / 2$ when $N$ is even. For notational convenience, we work with the case where $N$ is odd for our illustration.

Since the global sensitivity of the median is $\Delta^G = R$ as explained in Section \ref{dp:sensitivity}, it does not depend on the sample size. Hence, using the standard Laplace mechanism would be an ideal candidate to illustrate potential accuracy gains from sampling. However, as discussed in Section \ref{dp:illustrations}, due to the high global sensitivity of the median, this approach will hardly ever be useful in practice. Thus, we instead consider achieving $(\epsilon,\delta)$-DP for the population median by adding Laplace noise based on smooth sensitivity.

To create a privatized population median $\mathcal{A}(m_N)$, we consider its smooth sensitivity $\Delta^S_{\epsilon, \delta}(\bmY_N) = \textrm{max}_{k = 0, \cdots, N}\left(e^{-k\beta} \cdot \textrm{max}_{t = 0, \cdots, k+1} \left(y_{\frac{N + 1}{2} + t} - y_{\frac{N + 1}{2} + t - k - 1} \right)\right)$, where $\beta = \epsilon / \left( 2 \log(2 / \delta) \right)$. We use the randomized algorithm $\mathcal{A}(m_N) = m_N + \eta_N$ with $\eta_N \sim \textrm{Laplace}(0, 2\Delta^S_{\epsilon, \delta}(\bmY_N) / \epsilon)$. 

To create a privatized sample median, we draw a sample $\mathcal{S}(\bmY_N)$ of $n$ data points $y_1, \cdots, y_n$, 
and then apply the same procedure using the sample as input instead of the population, as well as the amplified privacy parameters $(\epsilon_n, \delta_n)$ from Theorem \ref{eq:amp_theorem}.

Unlike the global sensitivity, the smooth sensitivity is data dependent, which implies that we cannot analytically compare the variances of the noise component between the sample and the population as we have done in Section \ref{amplification:mean} for the mean. To illustrate this point, we note that
\begin{eqnarray*}
Var(\mathcal{A}(m_n))&=&E(Var(\mathcal{A}(m_n)|\mathcal{S}(Y_N)))+Var(E(\mathcal{A}(m_n)|\mathcal{S}(Y_N)))\\
&=&E(Var(m_n+\eta_n|\mathcal{S}(Y_N)))+Var(E(m_n+\eta_n|\mathcal{S}(Y_N)))\\
&=&E(Var(\eta_n|\mathcal{S}(Y_N)))+Var(m_n).
\end{eqnarray*}

The second component is just the variance of the sample median, which is easy to estimate at least approximately. However, the first component is more difficult, as the variance of the noise depends on the smooth sensitivity, which will change from sample to sample, making it difficult to compute the expectation over different samples.

Therefore, we will rely on simulations instead to illustrate that accuracy gains can be expected under certain regimes. We acknowledge that this implies that our findings will be specific to the distributional assumptions of the data in our simulations. This limitation seems to be unavoidable given that the amount of noise that is added will be data dependent. We still offer some general insights in which situations accuracy gains could be expected in Section \ref{sec:amplification:bi-modal}. To mimic a realistic scenario in which the median would typically be preferred over the mean, we assume that the variable of interest has a skewed distribution. Specifically, we generate a population of $N = 10,001$ records from a log-normal distribution with $\mu = 5$ and $\sigma = 0.5$. From this population we repeatedly ($T = 1,000$) draw simple random samples without replacement with varying sample sizes ($n\in \{1001,101\}$) and evaluate whether accuracy gains can be expected for different levels of protection ($\epsilon \in \{0.1,0.5,1\}$). We fix $\delta = 1/2N=4.9995 \times 10^{-5}$. For any of the six combinations of sample size and privacy parameters, the simulation consists of the following steps at each iteration $t=1,\ldots,T$:
\begin{enumerate}
    \item Generate a privatized median $\tilde{m}_N^{(t)}=\mathcal{A}(m_N)$ using the population and the target values of $\epsilon$ and $\delta$.
    \item
    Draw a simple random sample without replacement with sample size $n$ and calculate the sample median $m_n^{(t)}$. 
    \item
    Generate a privatized median $\tilde{m}_n^{(t)}=\mathcal{A}(m_n^{(t)})$ using the drawn sample and the amplified values $\epsilon_n$ and $\delta_n$.
\end{enumerate}

To evaluate the losses or gains in accuracy, we can compare the variability of $\tilde{m}_N$ and $\tilde{m}_n$ across the simulation runs.

\begin{figure}[t]
    \centering
    \includegraphics[width=0.3\textwidth]{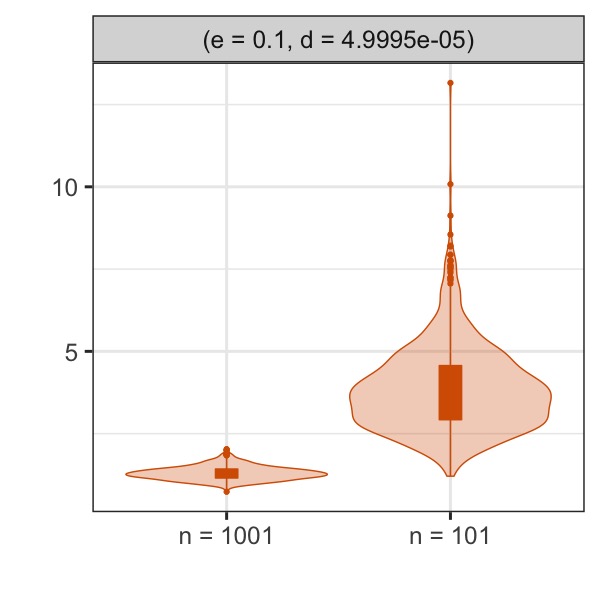}
    \includegraphics[width=0.3\textwidth]{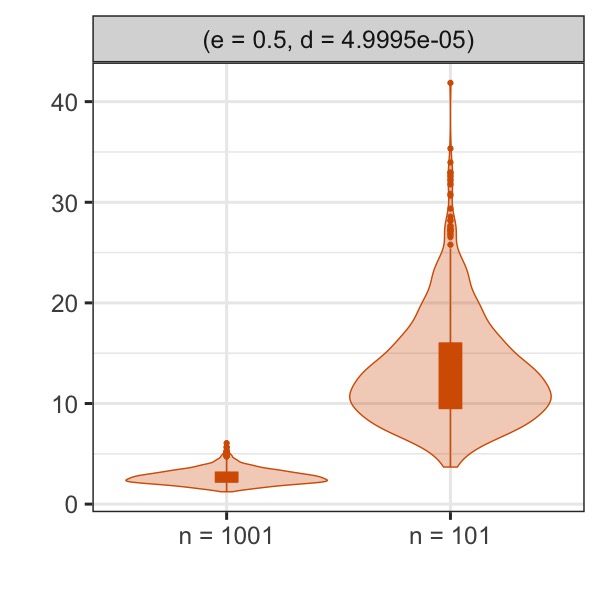}
    \includegraphics[width=0.3\textwidth]{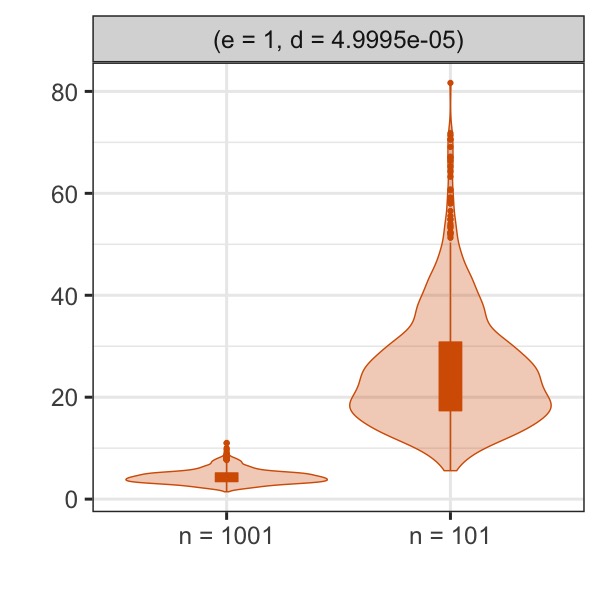}
    \caption{Violin plots of the ratio of the smooth sensitivity of the sample over the smooth sensitivity of the population for  $\epsilon = 0.1$ (left panel), $\epsilon = 0.5$ (middle panel), and $\epsilon = 1$ (right panel) and $n \in \{1001, 101\}$, based on 1,000 simulation runs. The parameter $\delta$ is fixed at $4.9995 \times 10^{-5}$ in all settings. 
    The population of size $N= 10,001$ is simulated from Lognormal(5, 0.5). 
    }
    \label{fig:Lognormal_all3_sensitivity}
\end{figure}

However, before presenting these comparisons, we empirically investigate the dependence between the smooth sensitivity of the median and the sample size. Figure \ref{fig:Lognormal_all3_sensitivity} depicts the ratio of the smooth sensitivity calculated from the $T = 1,000$ samples divided by the smooth sensitivity of the population. The sensitivity of the samples tends to be larger than the sensitivity in the population and it increases with smaller sample sizes and larger values of $\epsilon$ (note the different ranges of the $y$-axis for the three panels). To explain the inverse relationship between the smooth sensitivity and sample size, we note that the smooth sensitivity looks at the maximum of the weighted distance between neighboring records around the median, where the weights decrease with rank distance between the compared records. With smaller sample sizes, the distances between neighboring records will generally increase even if the data are generated from the same distribution. Hence we expect the smooth sensitivity to increase for smaller sampling rates. 

Increasing the value of $\epsilon$ implies that the weights decay more quickly, which generally means that the smooth sensitivity decreases for a fixed sample size with increasing values of $\epsilon$. However, the effective relative change of $\epsilon$ varies between the different datasets because of the amplification effects. While the underlying $\epsilon$ changes from 0.1 to 1, that is, it changes by a factor of 10, when going from the left panel to the right panel in Figure \ref{fig:Lognormal_all3_sensitivity}, the amplified $\epsilon_n$ for $n=101$ only changes from 2.43 to 5.14, that is, roughly by a factor of 2. Hence, the faster the sensitivity decreases, the smaller the amplification effects. Since the sensitivity will always decrease fastest in the population, we see that the ratio of the sensitivities as depicted in Figure \ref{fig:Lognormal_all3_sensitivity} will always increase with increasing values of $\epsilon$. Moreover, the increase will be more pronounced for smaller sampling rates.


However, we find that the increase in sensitivity is generally small, especially for small $\epsilon$. The median of the ratios for $\epsilon=0.1$ is 1.28 for $n=1,001$ and 3.72 for $n=101$ and even for $\epsilon=1$ the median of the ratios only increases to 4.24 and 23.30 respectively. Note that the global sensitivity for the mean discussed in Section \ref{amplification:mean} would increase by a factor of 10 for $n=1,001$ and by a factor of $100$ for $n=101$, irrespective of the selected level of $\epsilon$. 

\begin{figure}[t]
    \centering
    \includegraphics[width=0.3\textwidth]{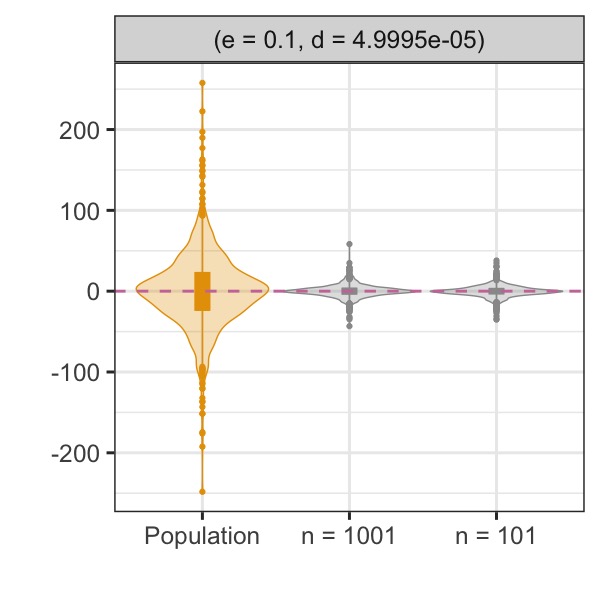}
    \includegraphics[width=0.3\textwidth]{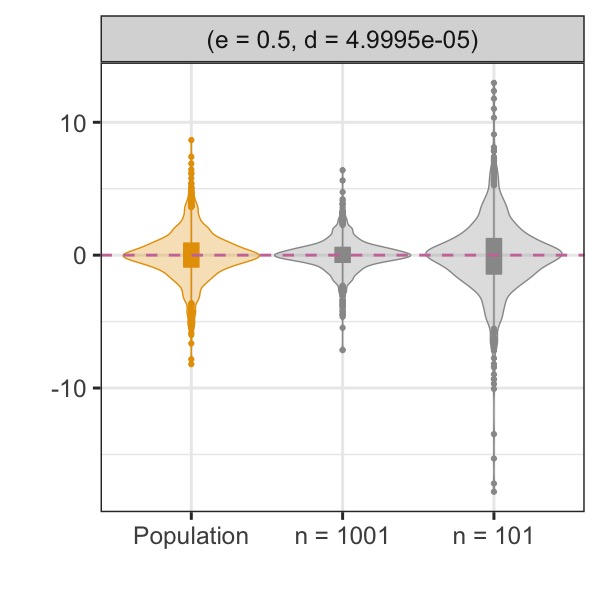}
    \includegraphics[width=0.3\textwidth]{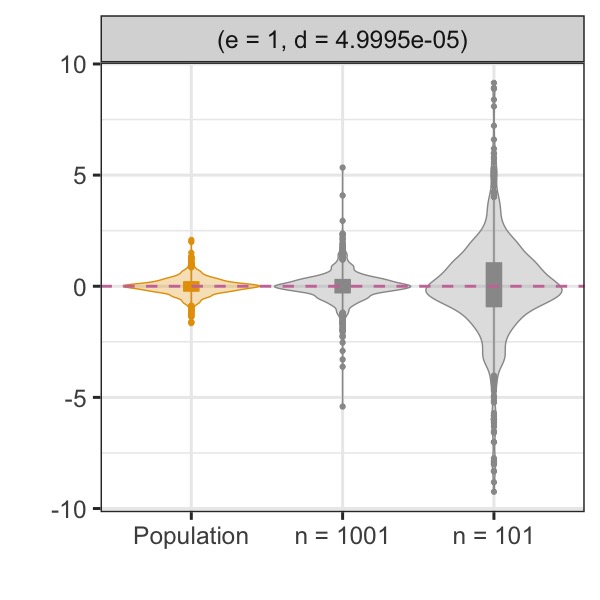}
    \caption{Violin plots showing the  distribution for the amount of noise added for privacy protection
    for $\epsilon = 0.1$ (left panel), $\epsilon = 0.5$ (middle panel), and $\epsilon = 1$ (right panel) and $n \in \{1001, 101\}$, based on 1,000 simulation runs. The parameter $\delta$ is fixed at $4.9995 \times 10^{-5}$ in all settings. The population of size $N= 10,001$ is simulated from Lognormal(5, 0.5).}
    \label{fig:Lognormal_all3_added_noise}
\end{figure}

As a first step to compare the variability of $\tilde{m}_N$ and $\tilde{m}_n$ across the simulation runs, Figure \ref{fig:Lognormal_all3_added_noise} contains violin plots showing the distribution for the amount of noise added for privacy protection across the simulation runs. For $\epsilon = 0.1$, we find a substantial reduction in the amount of noise when going from the population to the sample of size $n=1,001$. The noise is even further reduced for $n=101$.
However, these accuracy gains are lost once we increase the level of $\epsilon$. While the variance of the noise is reduced for all settings as expected if $\epsilon$ increases (note the different scales of the three panels), introducing a sampling step before adding the Laplace noise for privacy protection does no longer offer any advantages in terms of accuracy. Only for $\epsilon=0.5$ and a sample size of $n=1,001$ we still see a small reduction in the noise variance compared to the population. For all other scenarios, reducing the sample size will always increase the variance, indicating that the potential noise reduction from amplification is overcompensated by the increase in sensitivity.

\begin{figure}[t]
    \centering
    { \includegraphics[width=0.8\textwidth]{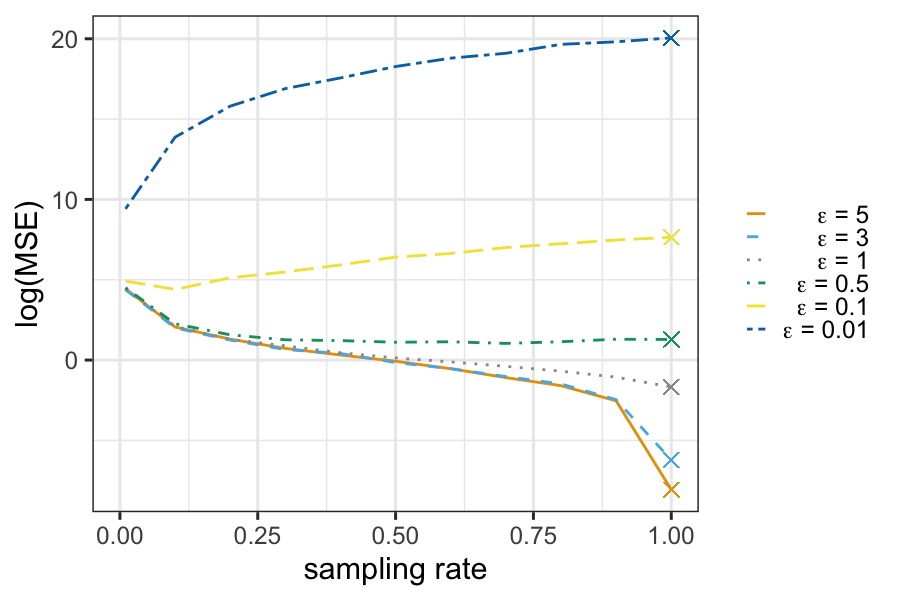}
    }\\
    \caption{Line plots showing log(MSE) as a function of sampling rate for noisy medians with $\epsilon \in \{0.01, 0.1, 0.5, 1, 3, 5\}$ and $\delta = 4.9995 \times 10^{-5}$, with sampling rates of \{$0.01, 0.1, 0.2, \cdots, 0.9$\}. The crosses at a sampling rate of 1 refer to the log(MSE) value of the population. The empirical log(MSE) is calculated based on 1,000 simulation runs. The population of size $N = 10,001$ is simulated from Lognormal(5, 0.5).}
    \label{fig:median_lognormal_master}
\end{figure}

These results have not yet accounted for the additional variance from sampling that will only increase with smaller sample sizes. Figure \ref{fig:median_lognormal_master} shows the empirical mean squared error (on the log-scale) of the final estimates $\tilde{m}_n$ as a function of the sampling rate for various values of $\epsilon$. We see that for small values of $\epsilon$ ($\epsilon\le 0.1$) accuracy gains from subsampling can be achieved. The gains are most pronounced for small sampling rates for $\epsilon\leq0.1$. However, for $\epsilon=0.5$, the accuracy starts decreasing again for sampling rates less than 10\%. For $\epsilon\ge 0.5$ no accuracy gains are achievable by subsampling the data.

However, there are specific data situations where accuracy gains can be achieved for larger values of $\epsilon$, as we will illustrate next.

\subsection{When sensitivity decreases with sample size}
\label{sec:amplification:bi-modal}


In Section \ref{amplification:variance} we claimed that in most scenarios the best we could hope for is that the sensitivity is independent of the sample size. This is certainly true for the global sensitivity. However, once we move to algorithms for which the sensitivity depends on the data, it is possible to construct scenarios in which the sensitivity actually decreases with sample size. We illustrate this phenomenon in this section using the smooth sensitivity of the median again. To ensure that the sensitivity in the sample will typically be smaller than in the population, we use a bi-modal distribution for which the true median lies in an area with low density between the two modes. Specifically, we generate our population of size $N = 10,001$ from a mixture of two scaled and shifted beta distributions where 5,001 records are drawn from the first mixture component and 5,000 records are drawn from the second mixture component. The R code used for simulating this bi-modal distribution is included in Appendix B.

\begin{figure}[t]
    \centering
    { \includegraphics[width=0.45\textwidth]{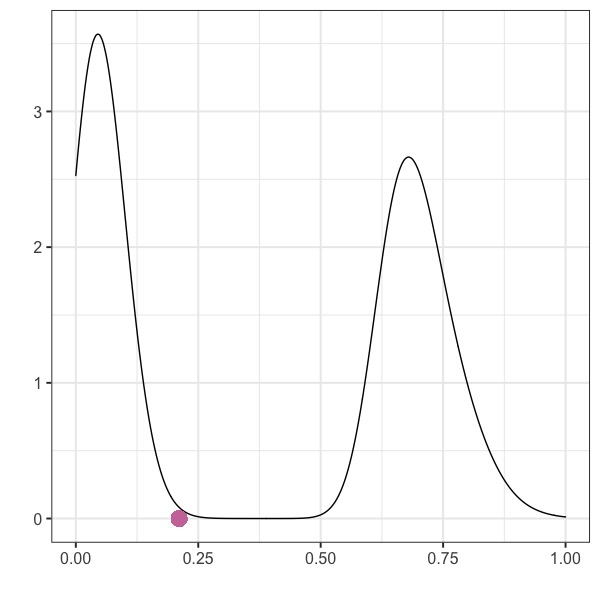}
    }
    \caption{Density plot of the simulated population of size $N=10,001$ based on a mixture of two scaled and shifted Beta distributions. The pink dot highlights the median of the distribution.} 
    \label{fig:Density_twobetas}
\end{figure}
The resulting distribution is plotted in Figure \ref{fig:Density_twobetas} with the value of the median, which is 0.21, highlighted as a dot. Note that the exact specification of the distributions is irrelevant for our illustrations as long as the two mixtures do not overlap and the median is equal to the largest value of the first mixture component (or the smallest value of the second mixture component). This will ensure that the smooth sensitivity will always be high in the population as the distance between the median and one of its neighbors will be large. 

\begin{figure}[t]
    \centering
    \includegraphics[width=0.3\textwidth]{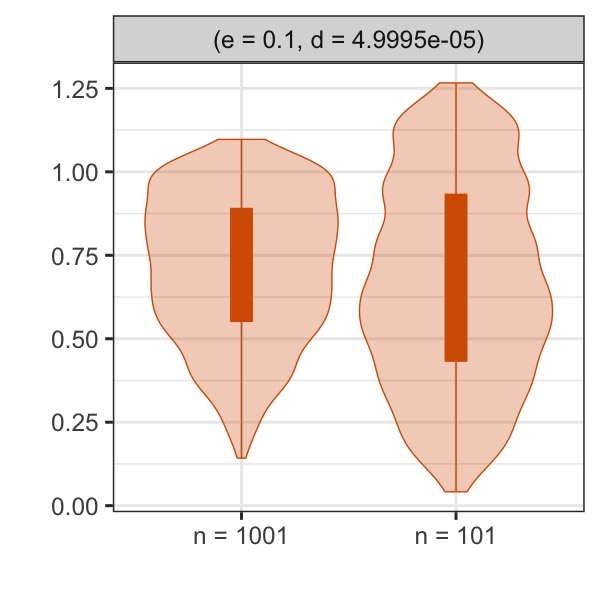}
    \includegraphics[width=0.3\textwidth]{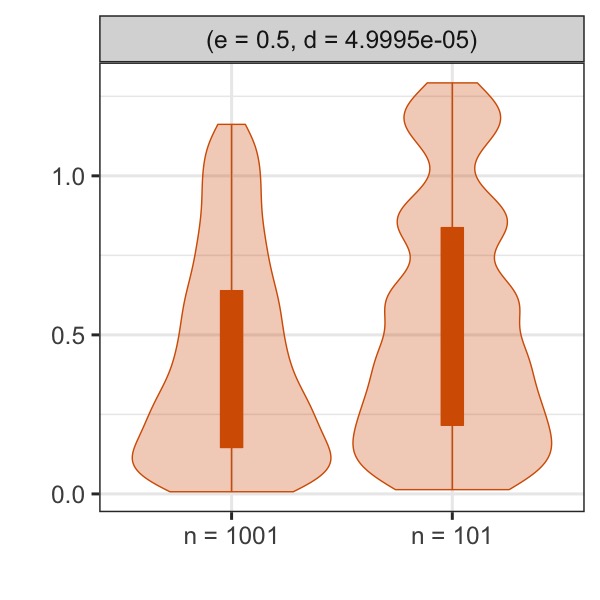}
    \includegraphics[width=0.3\textwidth]{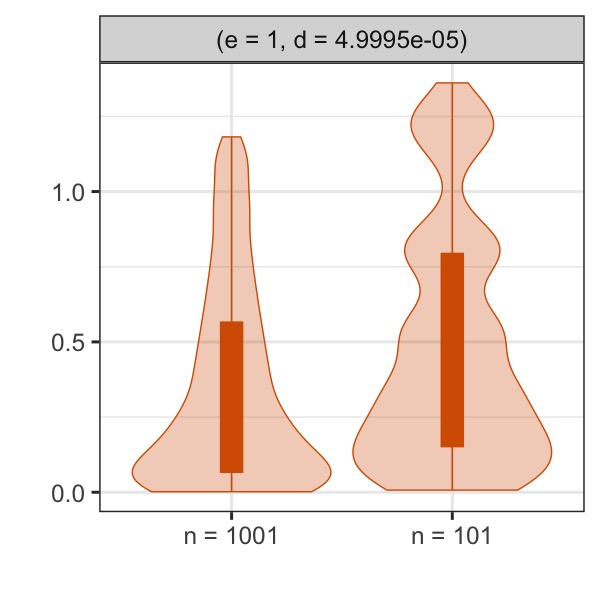}
    \caption{Violin plots of the ratio of the smooth sensitivity of the sample over the smooth sensitivity of the population for $\epsilon = 0.1$ (left panel), $\epsilon = 0.5$ (middle panel), and $\epsilon = 1$ (right panel) and $n \in \{1001, 101\}$, based on 1,000 simulation runs. The parameter $\delta$ is fixed at $4.9995 \times 10^{-5}$ in all settings. The population of size $N= 10,001$ is simulated from a mixture of two scaled and shifted Beta distributions.}
    \label{fig:twobetas_all3_ratios}
\end{figure}
\begin{figure}[!]
    \centering
    { \includegraphics[width=0.45\textwidth]{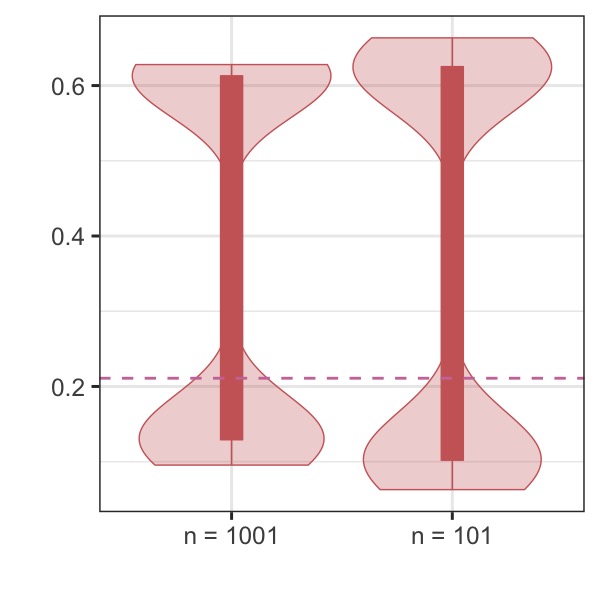}
    }\\
    \caption{Violin plots of the sample medians across 1,000 simulation runs. The population median is indicated by the dotted red line.}
    \label{fig:SampleMedian_twobetas_e1}
\end{figure}

Based on this population, we conduct the same simulation as in Section \ref{amplification:median}.
Results for the ratio of the sensitivity of the sample to the sensitivity in the population are depicted in Figure \ref{fig:twobetas_all3_ratios}. It is obvious that the sensitivity of the sample median is smaller than the sensitivity of the population median for most of the simulation runs.

Figure \ref{fig:SampleMedian_twobetas_e1} showing the distribution of the (unprivatized) sample medians helps to explain this somewhat unexpected result. 
The true median in the population is indicated by the dashed line. We find that the distribution of the sample medians is also bi-modal in this case with low density in the area of the true median. Comparing these distributions with the distribution of the data from Figure \ref{fig:Density_twobetas}, we note that most of the sample medians are in an area of the bi-modal distribution with much higher density than the area of the true median. Since the smooth sensitivity of the median increases with increasing sparsity of the data around the median
, the sensitivity of the population median is higher than the sensitivity for most of the sample medians.


Such a data scenario is ideal from an accuracy gain perspective, as both the reduced sensitivity and the amplified values of ($\epsilon_n, \delta_n$) will reduce the amount of noise that needs to be added to the sample data to achieve a target value for ($\epsilon, \delta$).

\begin{figure}[t]
    \centering
    \includegraphics[width=0.3\textwidth]{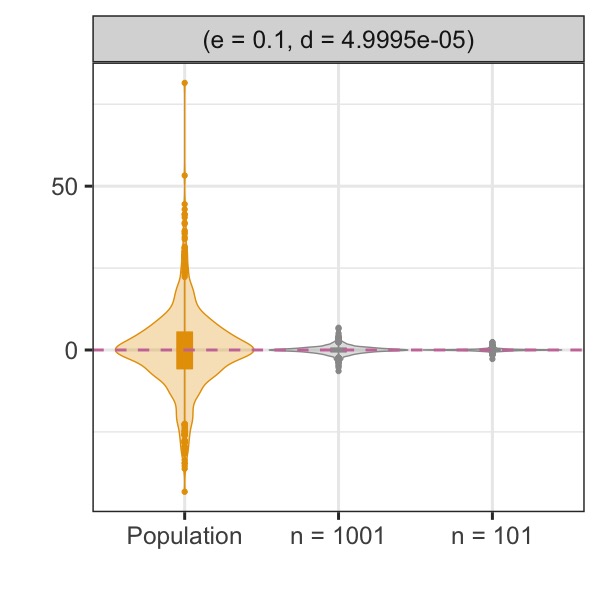}
    \includegraphics[width=0.3\textwidth]{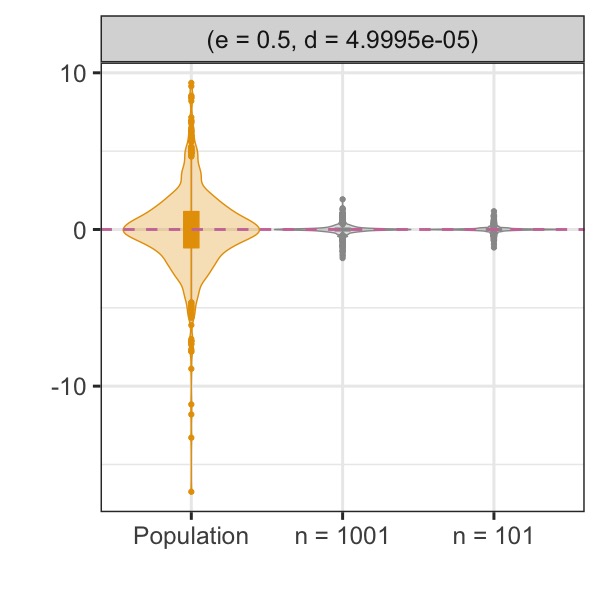}
    \includegraphics[width=0.3\textwidth]{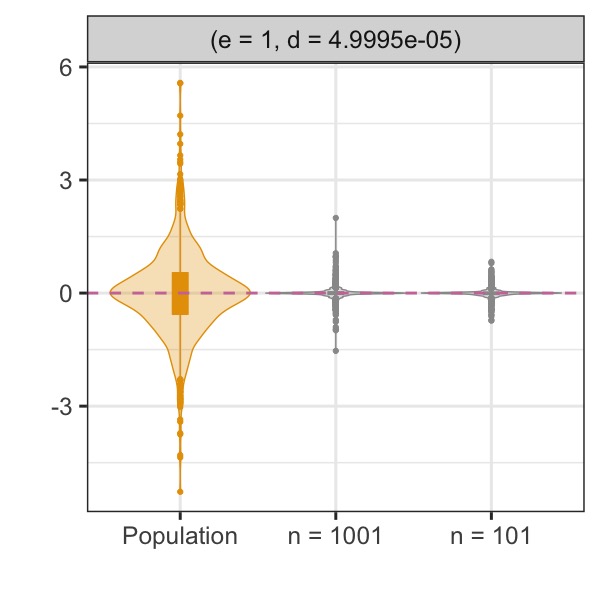}
    \caption{Violin plots showing the distribution for the amount of added noise for privacy protection for $\epsilon = 0.1$ (left panel), $\epsilon = 0.5$ (middle panel), and $\epsilon = 1$ (right panel) and $n \in \{1001, 101\}$, based on 1,000 simulation runs. The parameter $\delta$ is fixed at $4.9995 \times 10^{-5}$ in all settings. Population of size $N= 10,001$ simulated from a mixture of two scaled and shifted Beta distributions.}
    \label{fig:twobetas_all3_added_noise}
\end{figure}

\begin{figure}[t]
    \centering

    { \includegraphics[width=0.8\textwidth]{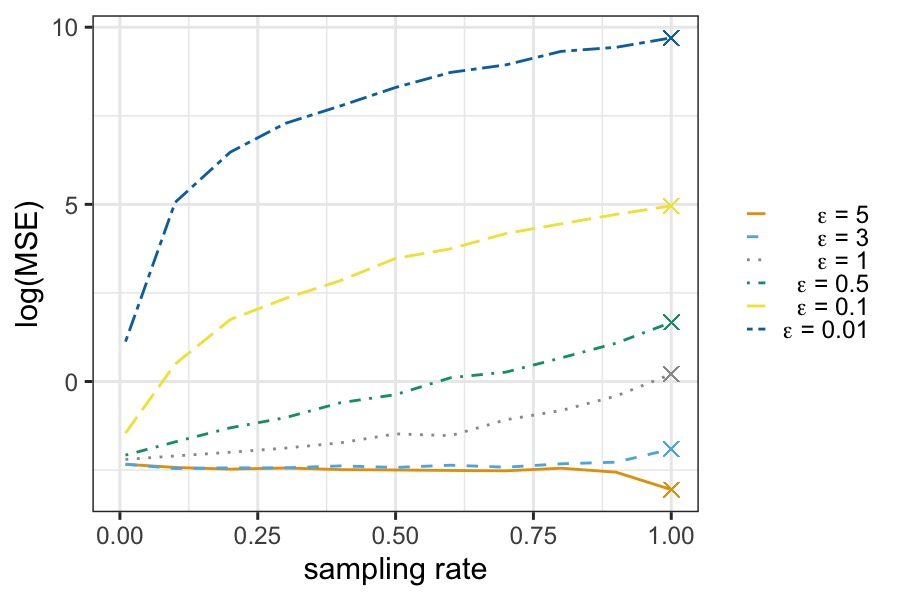}
    }\\
    \caption{Line plots showing log(MSE) as a function of sampling rate for noisy medians with $\epsilon \in \{0.01, 0.1, 0.5, 1, 3, 5\}$ and $\delta = 4.9995 \times 10^{-5}$, with sampling rates of \{$0.01, 0.1, 0.2, \cdots, 0.9$\}. The crosses at a sampling rate of 1 refer to the log(MSE) value of the population. The empirical log(MSE) is calculated based on 1,000 simulation runs. Population of size $N = 10,001$ simulated from a mixture of two scaled and shifted Beta distributions.}
    \label{fig:median_twobetas_master}
\end{figure}

This is confirmed in Figures \ref{fig:twobetas_all3_added_noise} and  \ref{fig:median_twobetas_master} 
which show the variability in the amount of noise and the mean squared error of the final estimates for the same scenarios considered in Section \ref{amplification:median} for the log-normal population.

Figure \ref{fig:median_twobetas_master} illustrates that gains can still be observed for $\epsilon$ values as large as 3. Moreover, the empirical log(MSE) values indicate that sampling is desirable even if we also account for the bias that is introduced by sampling in this scenario.

Admittedly, the substantial accuracy gains are only achievable as the simulation design ensures that the smooth sensitivity in the population is unusually large. Other differential privacy algorithms that are less sensitive to the sparsity of the data around the median  might produce more accurate results 
for this scenario. Therefore, it is unlikely
that similar gains would be achievable with these algorithms. The aim of this small illustration is to demonstrate that at least in principle, there could be scenarios in practice in which accuracy gains could be expected even for large values of $\epsilon$ that might be considered acceptable from an accuracy perspective. Some interesting questions for future research include whether such situations are likely to occur in practice, how a user would identify such a situation without spending privacy budget, and whether adding Laplace noise would still be a good protection strategy for such a scenario.  

\section{Concluding Remarks}
\label{conclusion}

Most research on differential privacy focuses on ensuring formal privacy guarantees for various information release scenarios. In this paper, we took a different perspective. We started from the assumption that the target level of protection has already been selected and that differential privacy algorithms exist that could be used to achieve this level of protection for the statistic to be released. Borrowing from the literature on privacy amplification through sampling, we asked the question whether it would be possible to improve the accuracy of the privatized estimates, if a sample is drawn first and the 
differentially private algorithm is run on the sample instead of the entire dataset. 

Beyond the obvious competing effects -- increasing sampling variance against decreasing variance from noise infusion because of privacy amplification -- we identified a third component that is critical when hoping for accuracy gains through sampling: the sensitivity, more specifically, the relationship between the sensitivity and the size of the database. As we illustrated using the Laplace mechanism for the mean, accuracy gains typically can not be expected if the sensitivity strongly depends on the size of the database. While the amplification effects imply that larger values of $\epsilon$ could be used to decrease the noise, this effect will often be compensated by the increase in sensitivity, resulting in increased level of noise. Since the sampling variance will also increase, very small values of $\epsilon$ would be required to see accuracy gains for this scenario. 

However, we also demonstrated that accuracy gains could be achieved in certain settings, if the sensitivity does not strongly depend on the sample size. Using smooth sensitivity coupled with Laplace noise to protect the median, we found that gains could be observed for values of ($\epsilon, \delta$) that might be considered realistic in practical applications.  
\begin{center}
\begin{table}[t]
    \centering
    \begin{tabular}{l l l l l }
    \hline \hline 
         Param. & Distribution & Sensitivity & DP def. & Conclusion  \\ \hline \hline\noalign{\smallskip}
    \multirow{3}{*}{Any} & \multirow{3}{*}{Any} & \multirow{3}{*}{Global} & \multirow{3}{*}{$\epsilon$} & No gains if\\
    &&&&$\textrm{Var}(g(\mathcal{S}(\bmY_N)))\geq$\\
    &&&&$2 (\Delta_N^{G})^2(\epsilon^{-2}-\epsilon_n^{-2})$\\ \noalign{\smallskip}\hline\noalign{\smallskip}  
         Mean & Any & Global & $\epsilon$ &No gains \\ \noalign{\smallskip}\hline\noalign{\smallskip}
        \multirow{3}{*}{Median} 
         & Lognormal(5, 0.5) & Smooth & $(\epsilon, \delta)$ &  Gains for $\epsilon \le 0.1$ \\[0.1cm]
         & Mixture of scaled and  &\multirow{2}{*}{Smooth} & \multirow{2}{*}{$(\epsilon, \delta)$} & \multirow{2}{*}{Gains for $\epsilon \leq 3$}\\
         &shifted Beta distributions \\\hline 
    \end{tabular}
    \caption{Summary of findings from the paper. The results based on global sensitivity assume that the Laplace mechanism is used. The $(\epsilon, \delta)-$DP in the table is with $\delta = 4.9995 \times 10^{-5}$.}
    \label{tab:results}
\end{table}
\end{center}

Table \ref{tab:results} summarizes the results from our simulation studies. We note that even if the results for the median are specific to the selected distributions, accuracy gains will generally be more likely if the data are sparse around the median, as it will be more likely that the smooth sensitivity will decrease when going from the population to the sample. Nevertheless, the research presented in this paper is limited in several dimensions. We only focus on mechanisms that rely on noise addition to protect the data and while we offer some general insights for such algorithms in Section \ref{amplification:variance}, we discuss amplification effects only for the mean and the median using only two differential privacy algorithms from the long list of mechanisms that could be considered to protect these two statistics. It would certainly be interesting, to study other popular mechanisms such as, for example, the Gaussian mechanism \citep{DworkRoth2014}, the exponential mechanism \citep{McSherryTalwar2007} or the Propose-Test-Release algorithm \citep{DworkLei2009STOC}. 
However, understanding the complex relationship between the selected privacy parameters, the sensitivity, and the resulting noise is beyond the scope of this introductory paper.

We also note that we only focus on the accuracy gains for a single statistic. In practice, statistical agencies will release a large number of statistics from their data. If gains could be expected for some of them but not for others, questions arise how to measure the overall gains to decide whether sampling strategies should be employed or not. Such a measure would not only have to account for different variances of the different estimates, it would also need to reflect that some of the estimates would be considered more important than others. Given that no accuracy gains can be expected in most of the scenarios that we considered in this paper, we feel that a discussion on how to optimally weight the accuracy gains and losses for different estimates when releasing multiple estimates seems premature. 
Furthermore, we emphasize that once multiple means are released from the same dataset, the Laplace mechanism might no longer be the best approach to protect the output \citep{steinke2016}. It would be interesting to study the privacy amplification of the two stage procedure proposed in \citet{steinke2016}, which was further developed in \citet{dagan2020} and \citet{ghazi2021}, to see whether accuracy gains from sampling might be possible under these regimes.

Another possible extension of our work would be to explore potential accuracy gains for other quantiles beyond the median. 
Our results for the median are not directly generalizable to other quantiles as both the smooth sensitivity as well as the sampling variance will change. Given that both quantities are data dependent, accuracy gains or losses will always be instance specific. As a general rule, accuracy gains will be more likely if the quantile of interest is in a low density area of the distribution as illustrated in Section \ref{sec:amplification:bi-modal}, as this will increase the chances of reducing the smooth sensitivity through sampling.

Another interesting area for future research would be to try to derive the optimal sampling rate from an accuracy perspective, that is, finding the sampling rate that minimizes the final variance of the privatized estimate. It must be noted, however, that since the final variance of many estimators depends on the variance in the population, this will typically not be possible without making further assumptions regarding the distribution of the data.  

Despite focusing on specific examples, this paper clearly illustrates that accuracy gains can only be expected if the sensitivity of the output does not depend too strongly on the size of the database. Thus, as a general recommendation, finding combinations of outputs and differential privacy algorithms for which the sensitivity does not depend on the data is key to achieve accuracy gains through (sub)sampling.



\bibliographystyle{natbib}
\bibliography{DPbib}

\begin{thebibliography}{}

\bibitem[Abadi \emph{et~al.}(2016)Abadi, Chu, Goodfellow, McMahan, Mironov,
  Talwar, and Zhang]{Abadi16}
Abadi, M., Chu, A., Goodfellow, I., McMahan, H.~B., Mironov, I., Talwar, K.,
  and Zhang, L. (2016).
\newblock Deep learning with differential privacy.
\newblock In \emph{Proceedings of the 2016 ACM SIGSAC Conference on Computer
  and Communications Security},  308--318.

\bibitem[Balle \emph{et~al.}(2020)Balle, Barthe, and Gaboardi]{Balle:2018}
Balle, B., Barthe, G., and Gaboardi, M. (2020).
\newblock Privacy profiles and amplification by subsampling.
\newblock \emph{Journal of Privacy and Confidentiality} \textbf{10}, 1.

\bibitem[Bun \emph{et~al.}(2020)Bun, Drechsler, Gaboardi, and
  McMillan]{bun2020}
Bun, M., Drechsler, J., Gaboardi, M., and McMillan, A. (2020).
\newblock Controlling privacy loss in survey sampling.
\newblock \emph{arXiv preprint arXiv:2007.12674} .

\bibitem[Bun \emph{et~al.}(2015)Bun, Nissim, Stemmer, and Vadhan]{BunNSV15}
Bun, M., Nissim, K., Stemmer, U., and Vadhan, S. (2015).
\newblock Differentially private release and learning of threshold functions.
\newblock In \emph{Proceedings of the 56th Annual IEEE Symposium on Foundations
  of Computer Science}, FOCS '15,  634--649, Washington, DC, USA. IEEE Computer
  Society.

\bibitem[Dagan and Kur(2020)]{dagan2020}
Dagan, Y. and Kur, G. (2020).
\newblock A bounded-noise mechanism for differential privacy.
\newblock \emph{arXiv preprint arXiv:2012.03817} .

\bibitem[Drechsler(2011)]{Drechsler2011book}
Drechsler, J. (2011).
\newblock \emph{Synthetic Datasets for Statistical Disclosure Control}.
\newblock Springer: New York.

\bibitem[Drechsler and Reiter(2010)]{drechsler2010}
Drechsler, J. and Reiter, J.~P. (2010).
\newblock Sampling with synthesis: A new approach for releasing public use
  census microdata.
\newblock \emph{Journal of the American Statistical Association} \textbf{105},
  492, 1347--1357.

\bibitem[Drechsler and Reiter(2012)]{drechsler2012}
Drechsler, J. and Reiter, J.~P. (2012).
\newblock Combining synthetic data with subsampling to create public use
  microdata files for large scale surveys.
\newblock \emph{Survey Methodology}  73--79.

\bibitem[Dwork(2008)]{dwork2008}
Dwork, C. (2008).
\newblock Differential privacy: A survey of results.
\newblock In \emph{International conference on theory and applications of
  models of computation},  1--19. Springer.

\bibitem[Dwork \emph{et~al.}(2006{a})Dwork, Kenthapadi, McSherry, Mironov, and
  Naor]{dwork2006our}
Dwork, C., Kenthapadi, K., McSherry, F., Mironov, I., and Naor, M. (2006{a}).
\newblock Our data, ourselves: Privacy via distributed noise generation.
\newblock In \emph{Annual International Conference on the Theory and
  Applications of Cryptographic Techniques},  486--503. Springer.

\bibitem[Dwork and Lei(2009)]{DworkLei2009STOC}
Dwork, C. and Lei, J. (2009).
\newblock Differential privacy and robust statistics.
\newblock In \emph{Proceedings of the 2009 International Association for
  Computing Machinery Symposium of Theory of Computing (STOC)}.

\bibitem[Dwork \emph{et~al.}(2006{b})Dwork, McSherry, Nissim, and
  Smith]{Dwork:2006:CNS:2180286.2180305}
Dwork, C., McSherry, F., Nissim, K., and Smith, A. (2006{b}).
\newblock Calibrating noise to sensitivity in private data analysis.
\newblock In \emph{Proceedings of the Third Conference on Theory of
  Cryptography}, TCC'06,  265--284, Berlin, Heidelberg. Springer-Verlag.

\bibitem[Dwork and Roth(2014)]{DworkRoth2014}
Dwork, C. and Roth, A. (2014).
\newblock \emph{The Algorithmic Foundations of Differential Privacy}.
\newblock Foundations and Trends in Theoretical Computer Science.

\bibitem[Ghazi \emph{et~al.}(2021)Ghazi, Kumar, and Manurangsi]{ghazi2021}
Ghazi, B., Kumar, R., and Manurangsi, P. (2021).
\newblock On avoiding the union bound when answering multiple differentially
  private queries.
\newblock In \emph{Conference on Learning Theory},  2133--2146. PMLR.

\bibitem[Kasiviswanathan \emph{et~al.}(2011)Kasiviswanathan, Lee, Nissim,
  Raskhodnikova, and Smith]{KasiviswanathanLNRS11}
Kasiviswanathan, S.~P., Lee, H.~K., Nissim, K., Raskhodnikova, S., and Smith,
  A. (2011).
\newblock What can we learn privately?
\newblock \emph{SIAM Journal on Computing} \textbf{40}, 3, 793--826.

\bibitem[Li \emph{et~al.}(2012)Li, Qardaji, and Su]{LiQardajiSu2011}
Li, N., Qardaji, W., and Su, D. (2012).
\newblock On sampling, anonymization, and differential privacy or,
  $k$-anonymization meets differential privacy.
\newblock In \emph{Proceedings of the 7th ACM Symposium on Information,
  Computer and Communications Security},  32--33.

\bibitem[McSherry and Talwar(2007)]{McSherryTalwar2007}
McSherry, M. and Talwar, K. (2007).
\newblock Mechanism design via differential privacy.
\newblock In \emph{Proceedings of the 48th Annual IEEE Symposium on Foundations
  of Computer Science},  94--103.

\bibitem[Nissim \emph{et~al.}(2007)Nissim, Raskhodnikova, and
  Smith]{NissimRaskhodnikovaSmith2007ACM}
Nissim, K., Raskhodnikova, S., and Smith, A. (2007).
\newblock Smooth sensitivity and sampling in private data analysis.
\newblock In \emph{Proceedings of the 39th Annual ACM Symposium on Theory of
  Computing},  75--83.

\bibitem[Steinke and Ullman(2016)]{steinke2016}
Steinke, T. and Ullman, J. (2016).
\newblock Between pure and approximate differential privacy.
\newblock \emph{Journal of Privacy and Confidentiality} \textbf{7}, 2.

\bibitem[Wang \emph{et~al.}(2019)Wang, Balle, and Kasiviswanathan]{WangBK19}
Wang, Y., Balle, B., and Kasiviswanathan, S.~P. (2019).
\newblock Subsampled renyi differential privacy and analytical moments
  accountant.
\newblock In \emph{The 22nd International Conference on Artificial Intelligence
  and Statistics, {AISTATS} 2019, 16-18 April 2019, Naha, Okinawa, Japan},
  vol.~89 of \emph{Proceedings of Machine Learning Research},  1226--1235.
  {PMLR}.

\bibitem[Wang \emph{et~al.}(2016)Wang, Lei, and Fienberg]{WangLF16b}
Wang, Y.-X., Lei, J., and Fienberg, S.~E. (2016).
\newblock Learning with differential privacy: Stability, learnability and the
  sufficiency and necessity of {ERM} principle.
\newblock \emph{J. Mach. Learn. Res.} \textbf{17}, 183:1--183:40.

\end{thebibliography}

\appendix


\section*{Appendix A: Proof of Theorem 1}
\label{App:Proof1}

\noindent {\bf Proof.}

The following proof holds for any original population $\bmY_N = \{y_i : i = 1,\ldots,N \}$ and a hypothetical population $\bmY_N' = \{y'_i : i = 1,\ldots,N \}$ such that the values of $\bmY_N$ and $\bmY_N'$ differ only in one record. Without loss of generality, we fix $k$ and suppose that $y'_i=y_i$ for all $i \neq k$ and $y'_k \neq y_k$. Let $\bmZ = (Z_1,\ldots,Z_N)$ denote the set of sample inclusion indicators of all population units, that is, $Z_i=1$ if unit $i$ is selected into the sample and $Z_i=0$ otherwise. Let $\bmz$ denote the realization of $\bmZ$ under the probabilistic sampling scheme. 
\begin{align}
 & \Pr[ \mathcal{A} ( g(\mathcal{S}(\bmY_N)) ) \le \omega ] = \sum_{\bm z} \Pr[ \mathcal{A} ( g(\mathcal{S}(\bmY_N)) ) \le \omega, \bmZ = \bmz ] \hspace{15em} \nonumber \\
 & =  \sum_{\{  \bm z : z_k = 1 \}} \Pr[ \mathcal{A} ( g(\mathcal{S}(\bmY_N)) ) \le \omega, \bmZ = \bmz ] 
+ \sum_{\{  \bm z : z_k = 0 \}} \Pr[ \mathcal{A} ( g(\mathcal{S}(\bmY_N)) ) \le \omega, \bmZ = \bmz ] \label{eq:proof_step1} \\
& \text{Because } \Pr[ \bmZ = \bmz ] = \frac{1}{{N \choose n}} \text{ for all } \bmz \text{ under simple random sampling without replacement}, \hspace{15em} \nonumber \\
& \text{ the first term of \eqref{eq:proof_step1} can be written as} \sum_{\{  \bm z : z_k = 1 \}} \Pr[ \mathcal{A} ( g(\mathcal{S}(\bmY_N)) ) \le \omega | \bmZ = \bmz ] \Pr[ \bmZ = \bmz ], \text{ where } \nonumber  \\
& \sum_{\{  \bm z : z_k = 1 \}} \Pr[ \mathcal{A} ( g(\mathcal{S}(\bmY_N)) ) \le \omega | \bmZ = \bmz ] \nonumber \\
& \le \min\left( \sum_{\{  \bm z' : z_k' = 1 \}} ( e^\epsilon  \Pr [ \mathcal{A} ( S(\bmY_N') ) \le \omega | \bmZ = \bmz' ] + \delta), \right. \hspace{15em} \nonumber \\
& \BBBB \BBBB \left. \frac{n}{N-n} \sum_{\{  \bm z' : z_k' = 0 \}} ( e^\epsilon  \Pr [ \mathcal{A} ( S(\bmY_N') ) \le \omega | \bmZ = \bmz' ] + \delta ) \right)  \label{eq:ineq_from_amplification}
\end{align}
\begin{align}
& = e^\epsilon \min \left( \ \sum_{\{  \bm z' : z_k' = 1 \}} \Pr [ \mathcal{A} ( S(\bmY_N') ) \le \omega | \bmZ = \bmz' ], \right. \hspace{15em} \nonumber \\
& \BBBB \BBBB \left. \ \frac{n}{N-n} \sum_{\{  \bm z' : z_k' = 0 \}} \Pr [ \mathcal{A} ( S(\bmY_N') ) \le \omega | \bmZ = \bmz' ] \right) 
\ + \ \delta {N-1 \choose n-1}. \nonumber 
\end{align}
The inequality in \eqref{eq:ineq_from_amplification} holds by comparing $\mathcal{S}(\bmY_N)$ to its neighbor $S(\bmY_N')$ for both cases where $y_k'$ is included in $S(\bmY_N')$ or not. 
When $\mathcal{S}(\bmY_N)$ includes $y_k$, i.e., $z_k=1$, we can find \emph{only one} neighbor $S(\bmY_N')$ that includes $y_k'$. For all ${N-1 \choose n-1}$ possible choices of $\bmz$ with $z_k = 1$, there exist the same number of possible choices of $\bmz'$ with $z'_k = 1$ that make $\mathcal{S}(\bmY_N)$ and $S(\bmY_N')$ differ by one element. Because we assume that $\mathcal{A}$ ensures $(\epsilon,\delta)$-DP, it holds that 
$$ \sum_{\{ \bm z:z_k=1 \}} \Pr \left[ \ \mathcal{A}(g(\mathcal{S}(\bmY_N))) \le \omega \ | \bmZ = \bmz \right] \le \sum_{\{ \bm z:z'_k=1 \}} ( e^\epsilon \Pr \left[ \ \mathcal{A}(S(g(\bmY_N'))) \le \omega \ | \bmZ = \bmz' \right] + \delta ). $$

\noindent For the second term in the minimum function in \eqref{eq:ineq_from_amplification}, we compare $\mathcal{S}(\bmY_N)$ including $y_k$ to $S(\bmY_N')$ not including $y_k'$. For each $\bmz$ with $z_k=1$, there exists a total of $N-n$ possible choices of $\bmz'$ with $z_k'=0$ that make $\mathcal{S}(\bmY_N)$ and $S(\bmY_N')$ differ by one element. For each $\bmz'$ with $z_k'=0$, there exists a total of $n$ possible choices of $\bmz$ with $z_k=1$ that make $\mathcal{S}(\bmY_N)$ and $S(\bmY_N')$ differ by one element. Therefore, with the $(\epsilon,\delta)$-DP mechanism $\mathcal{A}$, it holds that
{\footnotesize 
$$ (N-n) \sum_{\{ \bm z:z_k=1 \}} \Pr \left[ \ \mathcal{A}(g(\bmY_N)) \le \omega \ | \bmZ = \bmz \right] \le n \sum_{\{ \bm z':z'_k=0 \}} ( e^\epsilon \Pr \left[ \ \mathcal{A}(g(\bmY_N')) \le \omega \ | \bmZ = \bmz' \right] + \delta ).$$
}

Because $\Pr[ \bmZ = \bmz ] = 1 / {N \choose n}$ for all $\bmz$ under simple random sampling, the second term of \eqref{eq:proof_step1} can be written as 
$$ \sum_{\{  \bm z : z_k = 0 \}} \Pr[ \mathcal{A} ( g(\mathcal{S}(\bmY_N)) ) \le \omega | \bmZ = \bmz ] \Pr[ \bmZ = \bmz ] $$ 
where
$\sum_{\{  \bm z : z_k = 0 \}} \Pr[ \mathcal{A} ( g(\mathcal{S}(\bmY_N)) ) \le \omega | \bmZ = \bmz ] = \sum_{\{  \bm z' : z_k' = 0 \}} \Pr[ \mathcal{A} ( g(\mathcal{S}(\bmY_N)) ) \le \omega | \bmZ = \bmz' ]$. 

Now, let $A = \sum_{\{  \bm z' : z'_k = 1 \}} \Pr [ \mathcal{A} ( S(\bmY_N') ) \le \omega | \bmZ = \bmz' ]$ 
and 
$B = \sum_{\{  \bm z' : z_k' = 0 \}} \  \Pr [ \mathcal{A} ( S(\bmY_N') ) \le \omega | \bmZ = \bmz' ]$.
Then, Equation \eqref{eq:proof_step1} can be expressed as
\begin{align*}
& \Pr[ \mathcal{A} ( g(\mathcal{S}(\bmY_N)) ) \le \omega ] \le \left( 
e^\epsilon \min \left( \ A, \ \frac{n}{N-n} B \right)
+ \delta {N-1 \choose n-1} + B \right) \frac{1}{{N \choose n}} \\
& \ \le e^\epsilon \left( \left[ \frac{1}{e^\epsilon} + \frac{n}{N} \left( 1-\frac{1}{e^\epsilon} \right)  \right] A  + \left[ 1 - \frac{1}{e^\epsilon} - \frac{n}{N} \left( 1-\frac{1}{e^\epsilon} \right)  \right] \frac{n}{N-n}  B \right) \frac{1}{{N \choose n}} +  \frac{n}{N} \delta + B \frac{1}{{N \choose n}} \\
& = \left[ 1 + \frac{n}{N} \left( e^\epsilon - 1 \right) \right] (A+B) \frac{1}{{N \choose n}} + \frac{n}{N} \delta \\
& = \left[ 1 + \frac{n}{N} \left( e^\epsilon - 1 \right) \right] \sum_{\bm z} \Pr [ \mathcal{A} ( S(\bmY_N') ) \le \omega,  \bmZ = \bmz ] +  \frac{n}{N} \delta \\
& = \left[ 1 + \frac{n}{N} \left( e^\epsilon - 1 \right) \right] \Pr [ \mathcal{A} ( S(\bmY_N') ) \le \omega ] +  \frac{n}{N} \delta. 
\end{align*}
The second inequality holds because $\min(a,b) \le k a + (1-k) b$ for $0 < k < 1$. $\B_\square$





\section*{Appendix B: R code used to simulate the bi-modal distribution in Section \ref{sec:amplification:bi-modal}}
\label{App:code}

\begin{verbatim}
N <- 10001
set.seed(123)
a <- 2
b <- 10
y_N_orig <- c(rbeta((N+1)/2, a, b) / 2, rbeta((N-1)/2, a, b) + 1)
y_N <- (y_N_orig - min(y_N_orig)) / (max(y_N_orig) - min(y_N_orig))
\end{verbatim}

\end{document}